\documentclass[twocolumn,superscriptaddress]{revtex4-1} 
\usepackage{graphicx}
\usepackage{epstopdf}
\usepackage{amsmath}
\usepackage{amssymb}
\usepackage[colorlinks ,linkcolor=blue,anchorcolor=blue,citecolor=blue,urlcolor=blue]{hyperref}

\usepackage{color}
\usepackage{ulem}
\newcommand {\B}{\textcolor {blue}}

\renewcommand{\v}[1]{{\bf #1}}

\newcommand{\w}{{\omega}}
\def\eqa{\begin{eqnarray}}
\def\eea{\end{eqnarray}}

\newcommand{\Ref}[1]{Ref.~\cite{#1}}
\newcommand{\Eq}[1]{Eq.~(\ref{#1})}
\newcommand{\Fig}[1]{Fig.~\ref{#1}}

\newcommand{\eq}{\begin{equation}}
\newcommand{\ee}{\end{equation}}
\newcommand{\nn}{\nonumber\\}
\newcommand{\<}{\langle}
\renewcommand{\>}{\rangle}

\newcommand{\p}{\partial}
\renewcommand{\t}[1]{{\tilde #1}}
\newcommand{\ua}{\uparrow}
\newcommand{\da}{\downarrow}
\newcommand{\ra}{\rightarrow}

\newcommand{\al}{\alpha}
\newcommand{\bt}{\beta}

\newcommand{\la}{\lambda}
\newcommand{\La}{\Lambda}

\renewcommand{\dag}{\dagger}

\newcommand{\si}{\sigma}

\newcommand{\cC}{ {\cal C} }
\newcommand{\cD}{ {\cal D} }

\newcommand{\cH}{ {\cal H} }

\newcommand{\cP}{ {\cal P} }

\newcommand{\cU}{ {\cal U} }

\newcommand{\beginsupplement}{%
        \setcounter{table}{0}
        \renewcommand{\thetable}{S\arabic{table}}%
        \setcounter{figure}{0}
        \renewcommand{\thefigure}{S\arabic{figure}}%
        \setcounter{section}{0}
        \renewcommand{\thesection}{S\arabic{section}}
        \renewcommand{\theequation}{S\arabic{equation}}
     }

\begin{document}

\title{Phonon enhancement of electronic orders and negative isotope effect in the Hubbard-Holstein model on a square lattice}

\author{Da Wang}
\affiliation{National Laboratory of Solid State Microstructures $\&$ School of Physics, Nanjing
University, Nanjing, 210093, China}
\author{Wan-Sheng Wang}
\affiliation{National Laboratory of Solid State Microstructures $\&$ School of Physics, Nanjing
University, Nanjing, 210093, China}
\author{Qiang-Hua Wang}
\affiliation{National Laboratory of Solid State Microstructures $\&$ School of Physics, Nanjing
University, Nanjing, 210093, China}
\affiliation{Collaborative Innovation Center of Advanced Microstructures, Nanjing University, Nanjing 210093, China}

\begin{abstract}
Looking for superconductors with higher transition temperature requires a guiding principle. In conventional superconductors, electrons pair up into Cooper pairs via the retarded attraction mediated by electron-phonon coupling. Higher-frequency phonon (or smaller atomic mass) leads to higher superconducting transition temperature, known as the isotope effect. Furthermore, superconductivity is the only instability channel of the metallic normal state. In correlated systems, the above simple scenario could be easily violated. The strong local interaction is poorly screened, and this conspires with a featured Fermi surface to promote various competing electronic orders, such as spin-density-wave, charge-density-wave and unconventional superconductivity. On top of the various phases, the effect of electron-phonon coupling is an intriguing issue.  Using the functional renormalization group, here we investigated the interplay between the electron correlation and electron-phonon coupling in a prototype Hubbard-Holstein model on a square lattice. At half-filling, we found spin-density-wave and charge-density-wave phases and the transition between them, while no superconducting phase arises. Upon finite doping, d-wave/s-wave superconductivity emerges in proximity to spin-density-wave/charge-density-wave phases. Surprisingly, lower-frequency Holstein-phonons are either less destructive, or even beneficial, to the various phases, resulting in a negative isotope effect. We discuss the underlying mechanism behind and the implications of such anomalous effects.
\end{abstract}
\maketitle


According to the standard Bardeen-Cooper-Schrieffer (BCS) theory of superconductivity,\cite{Bardeen1957} the only possible instability of a metallic normal state, described by the Landau-Fermi liquid, is the Cooper pairing toward superconductivity (SC) upon an attractive interaction.~\cite{Shankar994} The electron-phonon coupling (EPC) can mediate a retarded attractive interaction between electrons. This has to withstand the repulsive Coulomb interaction. Fortunately in conventional metals, the long-range Coulomb interaction is well screened and can be effectively replaced by a pseudo-potential for quasi-particles below the energy scale of the Debye frequency $\omega_D$.~\cite{Migdal1958,*Eliashberg1960,Scalapino1966,*Mcmillan1968} The transition temperature $T_c$ increases (linearly at weak-coupling) with $\omega_D$, known as the isotope effect. This has been a guiding principle in the search of superconductors with higher $T_c$, provided that EPC is the pairing glue. However, the simple BCS scenario could break down in many ways. When the Fermi surface touches van Hove singularities, or is nested, density-waves in the spin or charge channel would be favorable. A well-known example is the Peierls instability toward the charge-density-wave (CDW) phase in one-dimensional (1D) electron systems with EPC alone.~\cite{Heeger1988} In correlated electrons systems, the local interactions are poorly screened, leaving the various orders, such as the spin-density-wave (SDW), CDW and unconventional SC, close competitors to each other, when the Fermi surface is featured with van Hove singularities and/or nesting. In the Mott limit, the strong local interaction leads to  the formation of local spin moments in the first place. The effect of EPC in such cases is an intriguing issue. For example, a long standing question is whether EPC plays a significant role for d-wave pairing in copper-based~\cite{Batlogg1987,*Franck1993,*Zech1994,Zhao1997,*Keller2003,*Gweon2004,*Lee2006,Alexandrov2003,*Muller2014} and $s_\pm$-wave pairing in iron-based high-temperature superconductors.~\cite{Liu2009,*Shirage2009} As a first step toward the issue, one considers theoretically a simplest model with local Holstein phonons and local Hubbard interactions, the so-called Hubbard-Holstein model (HHM).
Much effort has been devoted to understand the various orders and the metal-insulator transition in the HHM in 1D \cite{Takada2003,*Clay2005,*Hardikar2007,*Hohenadler2013} and infinite dimensions \cite{Sangiovanni2005,*Sangiovanni2006,*Werner2007,Bauer2010,*Murakami2013}.
In view of unconventional SC other than s-wave SC (s-SC), such as d-wave SC (d-SC) in layered materials, here we consider a HHM on a 2D square lattice. We handle the interplay between electron correlation and EPC by the singular-mode functional renormalization group (SM-FRG).~\cite{Wang2012,Xiang2012a} Our main findings are as follows. At half-filling, SDW and CDW competes, but no SC phase arises. Upon finite doping, d-SC/s-SC emerges in proximity to SDW/CDW phases. More interestingly, lower-frequency Holstein-phonons are either less destructive, or even beneficial, to the various phases, resulting in a negative isotope effect.

The 2D HHM is described by the Hamiltonian
\begin{eqnarray}
H&&=-t\sum_{\langle ij \rangle\sigma}(c_{i\sigma}^\dag c_{j\sigma}+{\rm h.c.}) -\mu\sum_{i\sigma}n_{i\sigma} + \w_D\sum_i b_i^\dag b_i\nn
&& +U \sum_{i}(n_{i\ua}-\frac{1}{2})(n_{i\da}-\frac{1}{2})+\eta\sum_{i\sigma} n_{i\sigma}(b_i^\dag+b_i),
\end{eqnarray}
where $t$ is the nearest-neighbor hopping, $\mu$ the chemical potential, $U$ the local Hubbard interaction, $\omega_D$ the Holstein phonon frequency, and $\eta = g/\sqrt{2M\w_D}$. Here $g$ is the EPC matrix element and $M$ is the mass of the vibrating ion. Henceforth we set $t=1$ as the unit of energy. The EPC leads to a retarded attraction $\Pi_\nu=-\la W \omega_D^2/(\omega_D^2+\nu^2)$, where $\nu$ is the Matsubara frequency, $W=8$ the electron bandwidth, and $\la = g^2/(M\w_D^2 W)$ an average EPC constant (which depends on the spring constant $K=M\w_D^2$ rather than on $\w_D$ independently).

We treat the correlation effect and EPC by the SM-FRG.~\cite{Wang2012,Xiang2012a} In a nutshell, the idea is to get momentum-resolved pseudo-potential $V_{1234}$, as in $(1/2)c_{1\si}^\dagger c_{2\si'}^\dagger V_{1234} c_{3\si'} c_{4\si}$, to act on low-energy fermionic degrees of freedom up to a cutoff energy scale $\Lambda$ (for Matsubara frequency in our case). Henceforth the numerical index labels momentum/position, and we leave implicit the momentum conservation/translation symmetry. Starting from the local $U$ at $\La=\infty$, $V$ can evolve, as $\La$ is lowered, to be nonlocal and even diverging, due to corrections from both $U$ and $\Pi$, to arbitrary orders and in all possible ways (see \B{Supplementary Materials}). To see the instability channel, we extract from $V$ and $\Pi$ the effective interactions in the general CDW/SDW/SC channels,
\eqa
&& [~V_{CDW}~]_{(14)(32)} = 2[~V+\Pi_0~]_{1234} - [~V+\Pi_\La~]_{1243},\nn
&& [~ V_{SDW} ~]_{(13)(42)} = - [~ V + \Pi_\La ~]_{1234},\nn
&& [~ V_{SC} ~]_{(12)(43)} = [~ V+\Pi_\La ~]_{1234}. \label{eq:VX}
\eea
The left-hand sides are understood as matrices with composite indices. Notice that $\Pi$ is local/flat in real/momentum space, and $\Pi_{\nu =0}$ enters $V_{CDW}$ because $\Pi$ is direct in the charge channel. Since they all originate from $V+\Pi$, $V_{CDW/SDW/SC}$ have overlaps but are naturally treated on equal footing. The divergence of the leading attractive (i.e., negative) eigenvalue of $V_{SC/SDW/CDW}$ decides the instability channel, the associated eigenfunction and collective momentum describe the order parameter, and the divergence scale $\La_c$ is representative of $T_c$.  More technical details can be found in \B{Supplementary Materials}.



Before embarking on full-wedge FRG results, we digress to gain qualitative insights first from an approximation to FRG. We keep the local part of $V$ only so that the FRG reduces to a simple RG. We focus on half filling, where the particle-hole symmetry enables us to solve $V$ analytically (see \B{Supplementary Materials}),
\eqa V+\Pi_0 \sim (U+\Pi_0)\exp\left[\frac{\al\la W}{\w_D}(1-\frac{2}{\pi}\tan^{-1}\frac{\La}{\w_D}) \right],\label{Eq:Anal}\eea where $\al$ is a constant of order unity.
We find $(V_{SC},V_{SDW},V_{CDW})=(V+\Pi_\La,-V-\Pi_\La,V+2\Pi_0-\Pi_\La)$ are bounded, but the behavior of $V$ still provides interesting implications.  (1) We observe that $V_{SC} > V_{CDW}$ for any $\la >0$, so SC is absent at half filling. In fact, even if $\la = 0$ the SC and CDW channels are exactly degenerate. (2) If $U+\Pi_0=U-\la W=0$, there is a fixed line $V=U$, on which $V_{SDW}=V_{CDW}<0$. This implies a phase boundary between CDW and SDW. (The local interactions are dispersionless, but the nesting vector $(\pi,\pi)$ decides the CDW/SDW wavevector.) (3) If $U-\la W>0$ (or $<0$), $V$ is driven more (or less) repulsive so that SDW (or CDW) can be enhanced by EPC; (4) A lower $\w_D$ leads to stronger enhancement of $|V+\Pi_0|$, implying that softer phonons are beneficial for CDW and SDW in the respective phase regimes, resulting in a negative isotope effect for both phases. This effect can also be understood from the quasi-particle point of view. By the Lang-Firsov approximation (see \B{Supplementary Materials}), the bandwidth is narrowed through polaron effect by a factor
\eqa z\sim \exp\left[-\frac{\la W}{2\w_D}\frac{1+e^{\bt\w_D}}{e^{\beta \w_D}-1}\right],\eea where $\bt=1/T$. On the other hand, $U\ra U-\la W$ after the Lang-Firsov transform. The interaction to bandwidth ratio becomes $r=(U/W-\la)/z$. This ratio is amplified by $1/z$ as long as $|U-\la W|\neq 0$,  implying a phase boundary $U-\la W=0$ between CDW and SDW, and more interestingly, the amplifying factor $1/z$ is in nice agreement with the exponential factor in \Eq{Eq:Anal}, provided that $\La\sim T\ll \w_D$ for the above $z$ to be applicable. Since we used the bare band as the input, it is amazing that our FRG could foresee the effect of polaronic band narrowing through the renormalization of the interaction. The negative isotope effect uncovered above is therefore exactly a manifestation of the fact that the polaronic effect is stronger for softer phonons.

\begin{figure}
\includegraphics[width=0.45\textwidth]{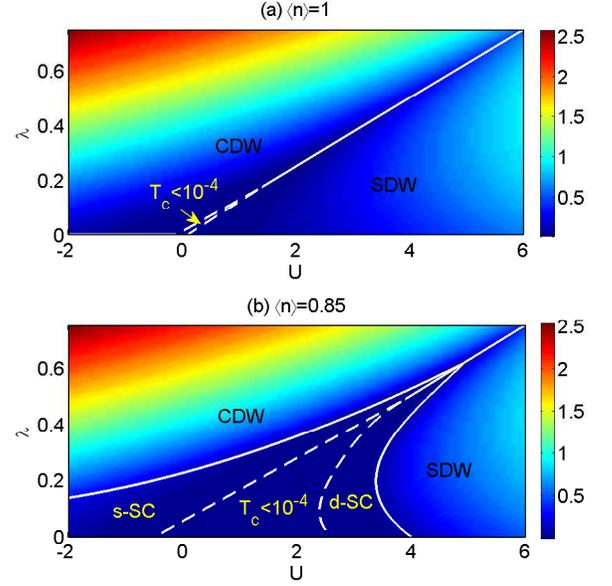}
\caption{Phase diagrams for (a) $\langle n\rangle = 1$ and (b) $\langle n\rangle = 0.85$. The color encodes the transition temperature $T_c$. The dashed lines enclose the regime in which $T_c<10^{-4}$. Here $\omega_D=0.5$.}
\label{fig:phase}
\end{figure}

We now turn to the full-wedge FRG results. \Fig{fig:phase} is the phase diagram for $\w_D=0.5$ for two filling levels. The color encodes $T_c$ versus $U$ and $\la$. The dashed lines enclose a regime in which $T_c < 10^{-4}$ beyond our interest. At half-filling $\<n\>=1$ in \Fig{fig:phase}(a), there is a phase boundary $U=\lambda W$ separating the CDW and SDW phases. In the CDW phase, $T_c$ is enhanced with increasing $\lambda$. In the SDW regime, $T_c$ exhibits a dome-shaped behavior along the $\lambda$-axis, implying that a weak $\lambda$ also enhances SDW. These behaviors are exactly what we discussed and understood in the above simple RG analysis. Moreover, the full FRG is able to capture general pairing channels. For example, in the SDW phase, $V_{SC}$ has a negative eigenvalue in the d-wave pairing channel, but it is always less diverging than that of $V_{SDW}$. This excludes d-SC at half filling. The phase-diagram is in full agreement with the quantum Monte Carlo (QMC) result~\cite{Nowadnick2012} on finite-size lattices, demonstrating the reliability of our FRG.

Away from half filling, our result for $\langle n \rangle =0.85$ is shown in \Fig{fig:phase}(b). Since the nesting is no longer at the Fermi level, the CDW (SDW) order is stabilized beyond a finite threshold $\lambda>\lambda_c$ ($U>U_c$). Below the phase boundary of CDW, we find s-SC is established. On the other hand, near the SDW phase boundary, d-SC emerges. The proximity between these phases is easily understood in view of the overlap in the SDW and SC channels, and is also known as a manifestation of pairing induced by SDW fluctuations.\cite{Scalapino1986,Bickers1989} What's more interesting here is the phase boundaries of both SDW and d-SC phases are curvy in the parameter space, implying that weak (strong) EPC enhances (suppresses) both SDW and d-SC. Given the behavior of SDW versus EPC we discussed above, however, the anomalous enhancement becomes natural in view of the overlap between SDW and d-SC channels. We notice that in an earlier FRG work,\cite{Honerkamp2007} the EPC (with Holstein-phonons) appears to suppress d-SC. We ascribe the difference to the dilute frequencies used for $\Pi$ in their case.

\begin{figure}
\includegraphics[width=0.5\textwidth]{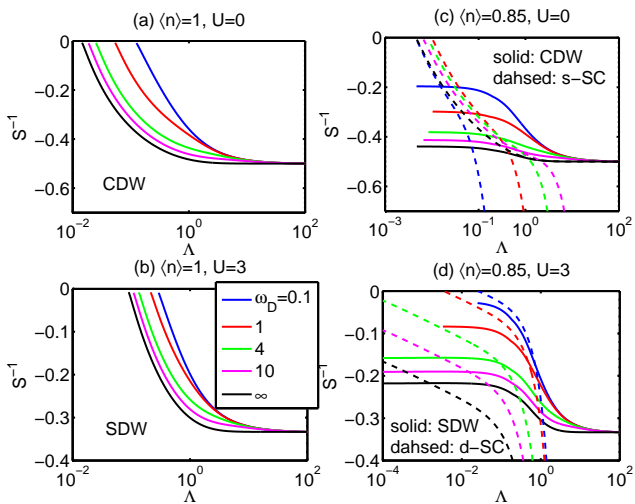}
\caption{Effects of EPC on the flow of leading eigenvalues $S$ (plotted as $1/S$ for clarity) of $V_{CDW/SDW/SC}$ at $\lambda=1/8$ for (a) $\<n\>=1$ and $U=0$, (b) $\<n\>=1$ and $U=3$, (c) $\<n\>=0.85$ and $U=0$, and (d) $\<n\>=0.85$ and $U=3$. The phonon frequency is indicated in the legend for all panels. For clarity, a channel is dropped if it's $|S|$ is too weak, and $S$ is multiplied by a factor of $10$ in the flow of SC channel in (d). }
\label{fig:flow}
\end{figure}

In order to have a closer view of the effects of EPC on the various orders, we plot some representative FRG flows of the leading eigenvalues $S$ of $V_{CDW/SDW/SC}$ in \Fig{fig:flow} for $\la=1/8$. Since we are looking for divergence, we drop out $\Pi_0$ and $\Pi_\La$ in \Eq{eq:VX} to concentrate on the flow of the projections of $V$ in the various channels. At half-filling with $U=0$/$3$ in \Fig{fig:flow}(a)/(b), the CDW/SDW channel diverges as $\Lambda$ is lowered. We have checked that for a pure negative-$U$ Hubbard model, equivalent to $\omega_D=\infty$ and $U=0$ in (a), the s-SC and CDW channels are exactly degenerate, satisfying the SO(4) $=$ SU(2) $\otimes$ SU(2) symmetry,\cite{Yang1990} where the excess pseudo-SU(2) arises from the particle-hole symmetry at half filling. However, a finite $\omega_D$ breaks the pseudo-SU(2) symmetry in favor of CDW,\cite{Noack1991} since $\Pi$ is a direct interaction in the charge channel.
For both CDW and SDW channels, $1/S$ is higher for lower $\w_D$, and so is $T_c$. This is just the negative isotope effect discussed earlier.

For $\<n\>=0.85$ in \Fig{fig:flow}(c)/(d), the CDW/SDW interaction flow at high $\La$ is similar to that in (a)/(b) for half-filling, since high energy quasi-particles are insensitive to the Fermi level. As $\La$ decreases further, however, low energy quasiparticles come into play, but the lack of nesting limits the phase space for low energy particle-hole excitations, so that the SDW/CDW channel eventually saturates. In contrary, there is no phase-space restriction for Cooper pairing, and upon an attractive pairing interaction, either already existing or induced via the overlap to CDW/SDW channels, the SC channel is boosted via the Cooper mechanism until it diverges. Not surprisingly, we find s-SC/d-SC in relation to the sub-leading CDW/SDW channel. More interestingly, the negative isotope effect for CDW and SDW clearly also acts on the proximiting SC, as is clear in \Fig{fig:flow}(c) and (d), and this is understood as from the channel overlap. The exception is the case of $\w_D=0.1$ in (c), which has the lowest $T_c$. In fact this is a case in the BCS limit, since $\la\ll 1$ and $\w_D\ll W$. We shall return to this point below.

\begin{figure}
\includegraphics[width=0.45\textwidth]{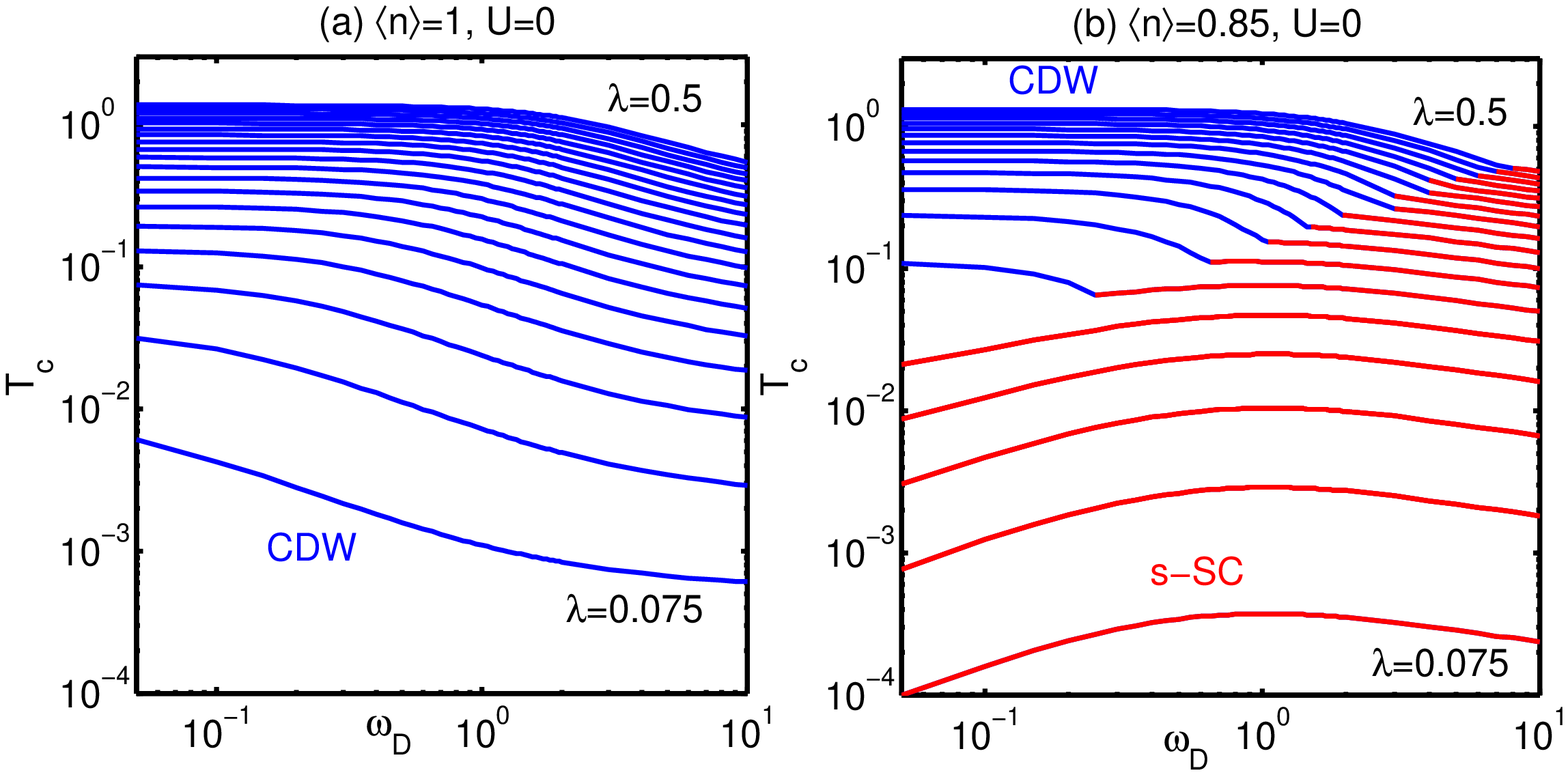}
\includegraphics[width=0.45\textwidth]{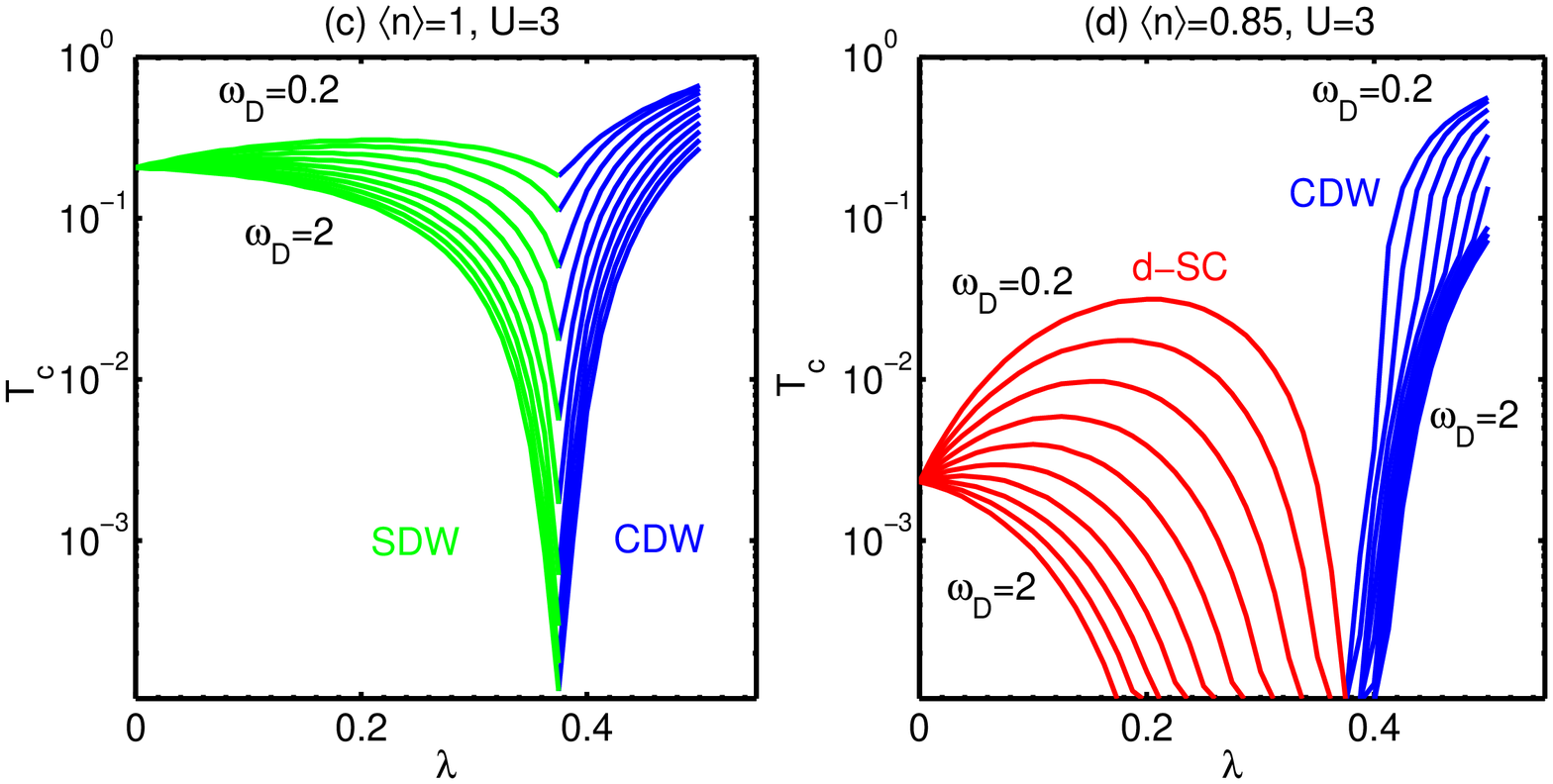}
\caption{$T_c$ versus $\lambda$ and $\omega_D$ with $U=0$ in (a) and (b), and $U=3$ in (c) and (d). The phases are denoted by both text and color. The solid lines are equally spaced by $\Delta\lambda=0.025$ in (a) and (b), and $\Delta\omega_D=0.2$ in (c) and (d). }
\label{fig:isotope}
\end{figure}

We now check more systematics for the effects of EPC on $T_c$ of the various phases. For a pure EPC system with $U=0$, the transition temperature $T_c$ is shown in \Fig{fig:isotope}(a) and (b). At half-filling in (a), only CDW phase is present, and $T_c$ clearly drops with increasing $\w_D$, for any $\la$. Such a negative isotope effect is discussed above, and is in agreement with the result judged from correlation functions measured by QMC on small clusters.\cite{Noack1991} For $\<n\>=0.85$ in (b), the CDW phase is realized for large $\la$, and $T_c$ follows the trend in (a) closely. For weaker $\la$, the system yields to the s-SC phase. In proximity to the CDW phase, we observe that $T_c$ for s-SC drops for larger $\w_D$. This is understood as caused by the weakening of CDW fluctuations, so that $T_c$ for s-SC eventually inherits a negative isotope effect. For even weaker $\la$, however, $T_c$ for s-SC increases with small $\w_D$, in a BCS fashion.
In fact, the FRG reproduces the exact BCS behavior $T_c\propto \w_D$ for $\la\ra 0$ and $\w_D\ll W$, if the CDW and SDW channels are too weak to affect the pairing channel (see \B{Supplementary Materials}). This is the case for $\la=1/8$ and $\w_D=0.1$ in Fig.~\ref{fig:flow}(c).
Therefore, the negative isotope effect for s-SC occurs only in proximity to the CDW phase.

For a correlated system with $U=3$, the transition temperature $T_c$ is shown in \Fig{fig:isotope}(c) and (d).  At half filling in (c), a large $\lambda$ drives SDW into CDW, with a phase transition at $\lambda_c=U/W$ independent of $\omega_D$. The transition temperature is always lower for larger $\w_D$, again a manifestation of the negative isotope effect. Moreover, we observe in (c) a slight enhancement of SDW by a weak $\la$ and small $\w_D$. This effect is qualitatively explained by \Eq{Eq:Anal}, and has been discussed previously. To our delight, the slight enhancement is consistent with the DCA result for $U=8$ in \Ref{Macridin2006}.
For the doped case, as $\<n\>=0.85$ in \Fig{fig:isotope}(d), the SDW phase yields to d-SC phase, and the CDW phase remains for large $\la$. Here $T_c$ is laterally higher for lower $\w_D$ for both CDW and d-SC in the respective regimes. A similar case was observed but only for the d-SC phase in Ref.\cite{Pao1998}. Moreover, in the d-SC regime, even though $T_c$ decreases with $\la$ for $\w_D>1$, it is lifted by a lower $\w_D$ for a given $\la$. Thus lower frequency phonons are at least less destructive to the d-SC. On the other hand, there is a marked enhancement of $T_c$ by a weak $\la$, up to $\la=0.2$ for $\w_D=0.2$, which we ascribe to the anomalous enhancement of SDW fluctuations as revealed in (c). The reason that the negative isotope effect is observed in the entire d-SC regime is because d-SC occurs only in proximity to the SDW phase.

\begin{figure}
\includegraphics[width=0.5\textwidth]{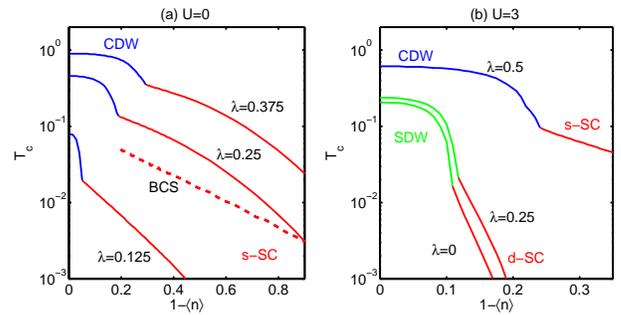}
\caption{Doping dependence of $T_c$ for $\w_D=0.5$ and $U=0$ (a) and $U=3$ (b). The phases are denoted by both text and color. The dashed line in (a) is a fit to the BCS theory (see the text for more details).}
\label{fig:doping}
\end{figure}

Finally, we consider the systematics in doping. We set $\w_D=0.5$ for illustration. For $U=0$, \Fig{fig:doping} (a) shows the CDW phase is present at low doping, and s-SC at higher doping. For stronger $\lambda$, a larger doping is needed to enter the s-SC phase. In both phases, $T_c$ decreases with doping, since the density of states $\rho$ at the Fermi level drops. We compare our result to the BCS formula (dashed line), $T_c^{BCS}=1.13\omega_D\exp(-1/\rho V_{BCS})$. \cite{Bardeen1957} We choose the value of $V_{BCS}$ so that $T_c^{BCS}$ matches our FRG result for $\lambda=0.25$ and a deep doping level $1-\< n\> =0.9$. We find $T_c>T_c^{BCS}$ approaching half filling. The enhancement follows from the effect of increasing CDW fluctuations, also favorable for s-wave pairing but missing in the simple BCS theory. For a nonzero $U=3$ in \Fig{fig:doping}(b), the CDW and s-SC phases are realized if $\la$ is sufficiently large (e.g., $\la=0.5$). For a weaker $\lambda=0.25$, d-SC sets in since the SDW fluctuations become stronger. Closer to half filling, the SDW phase eventually sets in. For both phases, $T_c$ is higher for $\la=0.25$ than that for $\la=0$, reconfirming the previous result that a weak EPC enhances SDW/d-SC if $\w_D$ is small.

In summary, we investigated the effects of EPC in a 2D HHM systematically by SM-FRG.
We found lower-frequency Holstein-phonons are beneficial to all of CDW, SDW, and s-SC/d-SC in proximity to CDW/SDW phases, resulting in a negative isotope effect. The qualitative mechanism is as follows. For CDW, low frequency phonons can be easily softened and adapt to the CDW order. Near the phase boundary of CDW, the enhanced CDW fluctuations are beneficial to s-SC. For SDW and d-SC, the enhancement can be effectively ascribed to polaronic band narrowing, which in turn blows up the correlation effect, favoring SDW and the d-SC in its proximity.

A few remarks are in order. (1) In a strict 2D system, academically there is no finite temperature SDW phase by the Mermin-Wagnar theorem,\cite{Mermin1966} and there is only algebraic SC order below the Kosterlitz-Thouless temperature.~\cite{Kosterlitz1973} In this regard, $T_c$ in our case should be understood as a crossover temperature in 2D, or the transition temperature in quasi-2D systems. (2) We should stress that FRG is perturbative in nature, and it works best in the itinerant picture up to moderate $U/W<1$ and $\la<1$ discussed in this paper. We expect application to cuprates in the overdoped region, upon necessary extension to more realistic phonon modes in cuprates. On the other hand, we expect application to the monolayer FeSe on the SrTiO3 substrate, where $T_c$ is found to be much higher than in the bulk.~\cite{Qing-Yan2012} The necessary extension is to account for the multi-orbitals in FeSe. Technically the FRG scheme in this paper improves the cutoff scheme in \Ref{Xiang2012}, and a further investigation of FeSe/SrTiO3 is in progress. (3) In the strong correlation limit, EPC might also enhance the $T_c$ of SDW~\cite{Macridin2006} as a result of self-localization of polarons in the presence of a sufficiently strong EPC.\cite{Zhong1992,Sakai1997,Huang2003,Mishchenko2004,Macridin2004,Macridin2006}. On the other hand, we notice that whether EPC would enhance d-SC in the strong correlation limit is under debate.~\cite{Huang2011,Macridin2012} Finally, there is even a proposal that EPC-driven bipolarons are necessary ingredients for high-$T_c$ SC in cuprates.~\cite{Alexandrov2003,*Muller2014}


\acknowledgements{The project was supported by NSFC (under grant No.10974086 and No.11023002) and the Ministry of Science and Technology of China (under grant No.2011CBA00108 and 2011CB922101). The numerical calculations were performed at the High Performance Computing Center of Nanjing University.}


\bibliography{hhfrg}

\begin{thebibliography}{52}%
\makeatletter
\providecommand \@ifxundefined [1]{%
 \@ifx{#1\undefined}
}%
\providecommand \@ifnum [1]{%
 \ifnum #1\expandafter \@firstoftwo
 \else \expandafter \@secondoftwo
 \fi
}%
\providecommand \@ifx [1]{%
 \ifx #1\expandafter \@firstoftwo
 \else \expandafter \@secondoftwo
 \fi
}%
\providecommand \natexlab [1]{#1}%
\providecommand \enquote  [1]{``#1''}%
\providecommand \bibnamefont  [1]{#1}%
\providecommand \bibfnamefont [1]{#1}%
\providecommand \citenamefont [1]{#1}%
\providecommand \href@noop [0]{\@secondoftwo}%
\providecommand \href [0]{\begingroup \@sanitize@url \@href}%
\providecommand \@href[1]{\@@startlink{#1}\@@href}%
\providecommand \@@href[1]{\endgroup#1\@@endlink}%
\providecommand \@sanitize@url [0]{\catcode `\\12\catcode `\$12\catcode
  `\&12\catcode `\#12\catcode `\^12\catcode `\_12\catcode `\%12\relax}%
\providecommand \@@startlink[1]{}%
\providecommand \@@endlink[0]{}%
\providecommand \url  [0]{\begingroup\@sanitize@url \@url }%
\providecommand \@url [1]{\endgroup\@href {#1}{\urlprefix }}%
\providecommand \urlprefix  [0]{URL }%
\providecommand \Eprint [0]{\href }%
\providecommand \doibase [0]{http://dx.doi.org/}%
\providecommand \selectlanguage [0]{\@gobble}%
\providecommand \bibinfo  [0]{\@secondoftwo}%
\providecommand \bibfield  [0]{\@secondoftwo}%
\providecommand \translation [1]{[#1]}%
\providecommand \BibitemOpen [0]{}%
\providecommand \bibitemStop [0]{}%
\providecommand \bibitemNoStop [0]{.\EOS\space}%
\providecommand \EOS [0]{\spacefactor3000\relax}%
\providecommand \BibitemShut  [1]{\csname bibitem#1\endcsname}%
\let\auto@bib@innerbib\@empty
\bibitem [{\citenamefont {Bardeen}\ \emph {et~al.}(1957)\citenamefont
  {Bardeen}, \citenamefont {Cooper},\ and\ \citenamefont
  {Schrieffer}}]{Bardeen1957}%
  \BibitemOpen
  \bibfield  {author} {\bibinfo {author} {\bibfnamefont {J.}~\bibnamefont
  {Bardeen}}, \bibinfo {author} {\bibfnamefont {L.~N.}\ \bibnamefont {Cooper}},
  \ and\ \bibinfo {author} {\bibfnamefont {J.~R.}\ \bibnamefont {Schrieffer}},\
  }\href {\doibase 10.1103/PhysRev.108.1175} {\bibfield  {journal} {\bibinfo
  {journal} {Phys. Rev.}\ }\textbf {\bibinfo {volume} {108}},\ \bibinfo {pages}
  {1175} (\bibinfo {year} {1957})}\BibitemShut {NoStop}%
\bibitem [{\citenamefont {Shankar}(1994)}]{Shankar994}%
  \BibitemOpen
  \bibfield  {author} {\bibinfo {author} {\bibfnamefont {R.}~\bibnamefont
  {Shankar}},\ }\href {\doibase 10.1103/RevModPhys.66.129} {\bibfield
  {journal} {\bibinfo  {journal} {Rev. Mod. Phys.}\ }\textbf {\bibinfo {volume}
  {66}},\ \bibinfo {pages} {129} (\bibinfo {year} {1994})}\BibitemShut
  {NoStop}%
\bibitem [{\citenamefont {Migdal}(1958)}]{Migdal1958}%
  \BibitemOpen
  \bibfield  {author} {\bibinfo {author} {\bibfnamefont {A.}~\bibnamefont
  {Migdal}},\ }\href@noop {} {\bibfield  {journal} {\bibinfo  {journal} {Sov.
  Phys. JETP}\ }\textbf {\bibinfo {volume} {7}},\ \bibinfo {pages} {996}
  (\bibinfo {year} {1958})}\BibitemShut {NoStop}%
\bibitem [{\citenamefont {Eliashberg}(1960)}]{Eliashberg1960}%
  \BibitemOpen
  \bibfield  {author} {\bibinfo {author} {\bibfnamefont {G.}~\bibnamefont
  {Eliashberg}},\ }\href@noop {} {\bibfield  {journal} {\bibinfo  {journal}
  {Sov. Phys. {JETP}}\ }\textbf {\bibinfo {volume} {11}} (\bibinfo {year}
  {1960})}\BibitemShut {NoStop}%
\bibitem [{\citenamefont {Scalapino}\ \emph {et~al.}(1966)\citenamefont
  {Scalapino}, \citenamefont {Schrieffer},\ and\ \citenamefont
  {Wilkins}}]{Scalapino1966}%
  \BibitemOpen
  \bibfield  {author} {\bibinfo {author} {\bibfnamefont {D.~J.}\ \bibnamefont
  {Scalapino}}, \bibinfo {author} {\bibfnamefont {J.~R.}\ \bibnamefont
  {Schrieffer}}, \ and\ \bibinfo {author} {\bibfnamefont {J.~W.}\ \bibnamefont
  {Wilkins}},\ }\href
  {http://journals.aps.org/pr/abstract/10.1103/PhysRev.148.263} {\bibfield
  {journal} {\bibinfo  {journal} {Phys. Rev.}\ }\textbf {\bibinfo {volume}
  {148}},\ \bibinfo {pages} {263} (\bibinfo {year} {1966})}\BibitemShut
  {NoStop}%
\bibitem [{\citenamefont {McMillan}(1968)}]{Mcmillan1968}%
  \BibitemOpen
  \bibfield  {author} {\bibinfo {author} {\bibfnamefont {W.~L.}\ \bibnamefont
  {McMillan}},\ }\href {\doibase 10.1103/PhysRev.167.331} {\bibfield  {journal}
  {\bibinfo  {journal} {Phys. Rev.}\ }\textbf {\bibinfo {volume} {167}},\
  \bibinfo {pages} {331} (\bibinfo {year} {1968})}\BibitemShut {NoStop}%
\bibitem [{\citenamefont {Heeger}\ \emph {et~al.}(1988)\citenamefont {Heeger},
  \citenamefont {Kivelson}, \citenamefont {Schrieffer},\ and\ \citenamefont
  {Su}}]{Heeger1988}%
  \BibitemOpen
  \bibfield  {author} {\bibinfo {author} {\bibfnamefont {A.~J.}\ \bibnamefont
  {Heeger}}, \bibinfo {author} {\bibfnamefont {S.}~\bibnamefont {Kivelson}},
  \bibinfo {author} {\bibfnamefont {J.~R.}\ \bibnamefont {Schrieffer}}, \ and\
  \bibinfo {author} {\bibfnamefont {W.~P.}\ \bibnamefont {Su}},\ }\href
  {\doibase 10.1103/RevModPhys.60.781} {\bibfield  {journal} {\bibinfo
  {journal} {Rev. Mod. Phys.}\ }\textbf {\bibinfo {volume} {60}},\ \bibinfo
  {pages} {781} (\bibinfo {year} {1988})}\BibitemShut {NoStop}%
\bibitem [{\citenamefont {Batlogg}\ \emph {et~al.}(1987)\citenamefont
  {Batlogg}, \citenamefont {Cava}, \citenamefont {Jayaraman}, \citenamefont
  {van Dover}, \citenamefont {Kourouklis}, \citenamefont {Sunshine},
  \citenamefont {Murphy}, \citenamefont {Rupp}, \citenamefont {Chen},
  \citenamefont {White}, \citenamefont {Short}, \citenamefont {Mujsce},\ and\
  \citenamefont {Rietman}}]{Batlogg1987}%
  \BibitemOpen
  \bibfield  {author} {\bibinfo {author} {\bibfnamefont {B.}~\bibnamefont
  {Batlogg}}, \bibinfo {author} {\bibfnamefont {R.~J.}\ \bibnamefont {Cava}},
  \bibinfo {author} {\bibfnamefont {A.}~\bibnamefont {Jayaraman}}, \bibinfo
  {author} {\bibfnamefont {R.~B.}\ \bibnamefont {van Dover}}, \bibinfo {author}
  {\bibfnamefont {G.~A.}\ \bibnamefont {Kourouklis}}, \bibinfo {author}
  {\bibfnamefont {S.}~\bibnamefont {Sunshine}}, \bibinfo {author}
  {\bibfnamefont {D.~W.}\ \bibnamefont {Murphy}}, \bibinfo {author}
  {\bibfnamefont {L.~W.}\ \bibnamefont {Rupp}}, \bibinfo {author}
  {\bibfnamefont {H.~S.}\ \bibnamefont {Chen}}, \bibinfo {author}
  {\bibfnamefont {A.}~\bibnamefont {White}}, \bibinfo {author} {\bibfnamefont
  {K.~T.}\ \bibnamefont {Short}}, \bibinfo {author} {\bibfnamefont {A.~M.}\
  \bibnamefont {Mujsce}}, \ and\ \bibinfo {author} {\bibfnamefont {E.~A.}\
  \bibnamefont {Rietman}},\ }\href {\doibase 10.1103/PhysRevLett.58.2333}
  {\bibfield  {journal} {\bibinfo  {journal} {Phys. Rev. Lett.}\ }\textbf
  {\bibinfo {volume} {58}},\ \bibinfo {pages} {2333} (\bibinfo {year}
  {1987})}\BibitemShut {NoStop}%
\bibitem [{\citenamefont {Franck}\ \emph {et~al.}(1993)\citenamefont {Franck},
  \citenamefont {Harker},\ and\ \citenamefont {Brewer}}]{Franck1993}%
  \BibitemOpen
  \bibfield  {author} {\bibinfo {author} {\bibfnamefont {J.}~\bibnamefont
  {Franck}}, \bibinfo {author} {\bibfnamefont {S.}~\bibnamefont {Harker}}, \
  and\ \bibinfo {author} {\bibfnamefont {J.}~\bibnamefont {Brewer}},\ }\href
  {\doibase 10.1103/PhysRevLett.71.283} {\bibfield  {journal} {\bibinfo
  {journal} {Phys. Rev. Lett.}\ }\textbf {\bibinfo {volume} {71}},\ \bibinfo
  {pages} {283} (\bibinfo {year} {1993})}\BibitemShut {NoStop}%
\bibitem [{\citenamefont {Zech}\ \emph {et~al.}(1994)\citenamefont {Zech},
  \citenamefont {Keller}, \citenamefont {Conder}, \citenamefont {Kaldis},
  \citenamefont {Liarokapis}, \citenamefont {Poulakis},\ and\ \citenamefont
  {M眉ller}}]{Zech1994}%
  \BibitemOpen
  \bibfield  {author} {\bibinfo {author} {\bibfnamefont {D.}~\bibnamefont
  {Zech}}, \bibinfo {author} {\bibfnamefont {H.}~\bibnamefont {Keller}},
  \bibinfo {author} {\bibfnamefont {K.}~\bibnamefont {Conder}}, \bibinfo
  {author} {\bibfnamefont {E.}~\bibnamefont {Kaldis}}, \bibinfo {author}
  {\bibfnamefont {E.}~\bibnamefont {Liarokapis}}, \bibinfo {author}
  {\bibfnamefont {N.}~\bibnamefont {Poulakis}}, \ and\ \bibinfo {author}
  {\bibfnamefont {K.~A.}\ \bibnamefont {M眉ller}},\ }\href {\doibase
  10.1038/371681a0} {\bibfield  {journal} {\bibinfo  {journal} {Nature}\
  }\textbf {\bibinfo {volume} {371}},\ \bibinfo {pages} {681} (\bibinfo {year}
  {1994})}\BibitemShut {NoStop}%
\bibitem [{\citenamefont {Zhao}\ \emph {et~al.}(1997)\citenamefont {Zhao},
  \citenamefont {Hunt}, \citenamefont {Keller},\ and\ \citenamefont
  {Muller}}]{Zhao1997}%
  \BibitemOpen
  \bibfield  {author} {\bibinfo {author} {\bibfnamefont {G.-m.}\ \bibnamefont
  {Zhao}}, \bibinfo {author} {\bibfnamefont {M.~B.}\ \bibnamefont {Hunt}},
  \bibinfo {author} {\bibfnamefont {H.}~\bibnamefont {Keller}}, \ and\ \bibinfo
  {author} {\bibfnamefont {K.~A.}\ \bibnamefont {Muller}},\ }\href
  {http://dx.doi.org/10.1038/385236a0} {\bibfield  {journal} {\bibinfo
  {journal} {Nature}\ }\textbf {\bibinfo {volume} {385}},\ \bibinfo {pages}
  {236} (\bibinfo {year} {1997})}\BibitemShut {NoStop}%
\bibitem [{\citenamefont {Keller}(2003)}]{Keller2003}%
  \BibitemOpen
  \bibfield  {author} {\bibinfo {author} {\bibfnamefont {H.}~\bibnamefont
  {Keller}},\ }\href {\doibase http://dx.doi.org/10.1016/S0921-4526(02)01632-0}
  {\bibfield  {journal} {\bibinfo  {journal} {Physica B: Condensed Matter}\
  }\textbf {\bibinfo {volume} {326}},\ \bibinfo {pages} {283 } (\bibinfo {year}
  {2003})}\BibitemShut {NoStop}%
\bibitem [{\citenamefont {Gweon}\ \emph {et~al.}(2004)\citenamefont {Gweon},
  \citenamefont {Sasagawa}, \citenamefont {Zhou}, \citenamefont {Graf},
  \citenamefont {Takagi}, \citenamefont {Lee},\ and\ \citenamefont
  {Lanzara}}]{Gweon2004}%
  \BibitemOpen
  \bibfield  {author} {\bibinfo {author} {\bibfnamefont {G.-H.}\ \bibnamefont
  {Gweon}}, \bibinfo {author} {\bibfnamefont {T.}~\bibnamefont {Sasagawa}},
  \bibinfo {author} {\bibfnamefont {S.~Y.}\ \bibnamefont {Zhou}}, \bibinfo
  {author} {\bibfnamefont {J.}~\bibnamefont {Graf}}, \bibinfo {author}
  {\bibfnamefont {H.}~\bibnamefont {Takagi}}, \bibinfo {author} {\bibfnamefont
  {D.-H.}\ \bibnamefont {Lee}}, \ and\ \bibinfo {author} {\bibfnamefont
  {A.}~\bibnamefont {Lanzara}},\ }\href {\doibase 10.1038/nature02731}
  {\bibfield  {journal} {\bibinfo  {journal} {Nature}\ }\textbf {\bibinfo
  {volume} {430}},\ \bibinfo {pages} {187} (\bibinfo {year}
  {2004})}\BibitemShut {NoStop}%
\bibitem [{\citenamefont {Lee}\ \emph {et~al.}(2006)\citenamefont {Lee},
  \citenamefont {Fujita}, \citenamefont {McElroy}, \citenamefont {Slezak},
  \citenamefont {Wang}, \citenamefont {Aiura}, \citenamefont {Bando},
  \citenamefont {Ishikado}, \citenamefont {Masui}, \citenamefont {Zhu},
  \citenamefont {Balatsky}, \citenamefont {Eisaki}, \citenamefont {Uchida},\
  and\ \citenamefont {Davis}}]{Lee2006}%
  \BibitemOpen
  \bibfield  {author} {\bibinfo {author} {\bibfnamefont {J.}~\bibnamefont
  {Lee}}, \bibinfo {author} {\bibfnamefont {K.}~\bibnamefont {Fujita}},
  \bibinfo {author} {\bibfnamefont {K.}~\bibnamefont {McElroy}}, \bibinfo
  {author} {\bibfnamefont {J.~A.}\ \bibnamefont {Slezak}}, \bibinfo {author}
  {\bibfnamefont {M.}~\bibnamefont {Wang}}, \bibinfo {author} {\bibfnamefont
  {Y.}~\bibnamefont {Aiura}}, \bibinfo {author} {\bibfnamefont
  {H.}~\bibnamefont {Bando}}, \bibinfo {author} {\bibfnamefont
  {M.}~\bibnamefont {Ishikado}}, \bibinfo {author} {\bibfnamefont
  {T.}~\bibnamefont {Masui}}, \bibinfo {author} {\bibfnamefont {J.-X.}\
  \bibnamefont {Zhu}}, \bibinfo {author} {\bibfnamefont {A.~V.}\ \bibnamefont
  {Balatsky}}, \bibinfo {author} {\bibfnamefont {H.}~\bibnamefont {Eisaki}},
  \bibinfo {author} {\bibfnamefont {S.}~\bibnamefont {Uchida}}, \ and\ \bibinfo
  {author} {\bibfnamefont {J.~C.}\ \bibnamefont {Davis}},\ }\href {\doibase
  10.1038/nature04973} {\bibfield  {journal} {\bibinfo  {journal} {Nature}\
  }\textbf {\bibinfo {volume} {442}},\ \bibinfo {pages} {546} (\bibinfo {year}
  {2006})}\BibitemShut {NoStop}%
\bibitem [{\citenamefont {Alexandrov}(2003)}]{Alexandrov2003}%
  \BibitemOpen
  \bibfield  {author} {\bibinfo {author} {\bibfnamefont {A.~S.}\ \bibnamefont
  {Alexandrov}},\ }\href@noop {} {\bibfield  {journal} {\bibinfo  {journal}
  {Theory of superconductivity: from weak to strong coupling (CRC Press)}\ }
  (\bibinfo {year} {2003})}\BibitemShut {NoStop}%
\bibitem [{\citenamefont {M\"uller}(2014)}]{Muller2014}%
  \BibitemOpen
  \bibfield  {author} {\bibinfo {author} {\bibfnamefont {K.~A.}\ \bibnamefont
  {M\"uller}},\ }\href {\doibase 10.1007/s10948-014-2751-5} {\bibfield
  {journal} {\bibinfo  {journal} {J. Supercond. Nov. Magn.}\ }\textbf {\bibinfo
  {volume} {27}},\ \bibinfo {pages} {2163} (\bibinfo {year}
  {2014})}\BibitemShut {NoStop}%
\bibitem [{\citenamefont {Liu}\ \emph {et~al.}(2009)\citenamefont {Liu},
  \citenamefont {Wu}, \citenamefont {Wu}, \citenamefont {Chen}, \citenamefont
  {Wang}, \citenamefont {Xie}, \citenamefont {Ying}, \citenamefont {Yan},
  \citenamefont {Li}, \citenamefont {Shi}, \citenamefont {Chu}, \citenamefont
  {Wu},\ and\ \citenamefont {Chen}}]{Liu2009}%
  \BibitemOpen
  \bibfield  {author} {\bibinfo {author} {\bibfnamefont {R.~H.}\ \bibnamefont
  {Liu}}, \bibinfo {author} {\bibfnamefont {T.}~\bibnamefont {Wu}}, \bibinfo
  {author} {\bibfnamefont {G.}~\bibnamefont {Wu}}, \bibinfo {author}
  {\bibfnamefont {H.}~\bibnamefont {Chen}}, \bibinfo {author} {\bibfnamefont
  {X.~F.}\ \bibnamefont {Wang}}, \bibinfo {author} {\bibfnamefont {Y.~L.}\
  \bibnamefont {Xie}}, \bibinfo {author} {\bibfnamefont {J.~J.}\ \bibnamefont
  {Ying}}, \bibinfo {author} {\bibfnamefont {Y.~J.}\ \bibnamefont {Yan}},
  \bibinfo {author} {\bibfnamefont {Q.~J.}\ \bibnamefont {Li}}, \bibinfo
  {author} {\bibfnamefont {B.~C.}\ \bibnamefont {Shi}}, \bibinfo {author}
  {\bibfnamefont {W.~S.}\ \bibnamefont {Chu}}, \bibinfo {author} {\bibfnamefont
  {Z.~Y.}\ \bibnamefont {Wu}}, \ and\ \bibinfo {author} {\bibfnamefont {X.~H.}\
  \bibnamefont {Chen}},\ }\href {\doibase 10.1038/nature07981} {\bibfield
  {journal} {\bibinfo  {journal} {Nature}\ }\textbf {\bibinfo {volume} {459}},\
  \bibinfo {pages} {64} (\bibinfo {year} {2009})}\BibitemShut {NoStop}%
\bibitem [{\citenamefont {Shirage}\ \emph {et~al.}(2009)\citenamefont
  {Shirage}, \citenamefont {Kihou}, \citenamefont {Miyazawa}, \citenamefont
  {Lee}, \citenamefont {Kito}, \citenamefont {Eisaki}, \citenamefont
  {Yanagisawa}, \citenamefont {Tanaka},\ and\ \citenamefont
  {Iyo}}]{Shirage2009}%
  \BibitemOpen
  \bibfield  {author} {\bibinfo {author} {\bibfnamefont {P.~M.}\ \bibnamefont
  {Shirage}}, \bibinfo {author} {\bibfnamefont {K.}~\bibnamefont {Kihou}},
  \bibinfo {author} {\bibfnamefont {K.}~\bibnamefont {Miyazawa}}, \bibinfo
  {author} {\bibfnamefont {C.-H.}\ \bibnamefont {Lee}}, \bibinfo {author}
  {\bibfnamefont {H.}~\bibnamefont {Kito}}, \bibinfo {author} {\bibfnamefont
  {H.}~\bibnamefont {Eisaki}}, \bibinfo {author} {\bibfnamefont
  {T.}~\bibnamefont {Yanagisawa}}, \bibinfo {author} {\bibfnamefont
  {Y.}~\bibnamefont {Tanaka}}, \ and\ \bibinfo {author} {\bibfnamefont
  {A.}~\bibnamefont {Iyo}},\ }\href {\doibase 10.1103/PhysRevLett.103.257003}
  {\bibfield  {journal} {\bibinfo  {journal} {Phys. Rev. Lett.}\ }\textbf
  {\bibinfo {volume} {103}},\ \bibinfo {pages} {257003} (\bibinfo {year}
  {2009})}\BibitemShut {NoStop}%
\bibitem [{\citenamefont {Takada}\ and\ \citenamefont
  {Chatterjee}(2003)}]{Takada2003}%
  \BibitemOpen
  \bibfield  {author} {\bibinfo {author} {\bibfnamefont {Y.}~\bibnamefont
  {Takada}}\ and\ \bibinfo {author} {\bibfnamefont {A.}~\bibnamefont
  {Chatterjee}},\ }\href {\doibase 10.1103/PhysRevB.67.081102} {\bibfield
  {journal} {\bibinfo  {journal} {Phys. Rev. B}\ }\textbf {\bibinfo {volume}
  {67}},\ \bibinfo {pages} {081102} (\bibinfo {year} {2003})}\BibitemShut
  {NoStop}%
\bibitem [{\citenamefont {Clay}\ and\ \citenamefont
  {Hardikar}(2005)}]{Clay2005}%
  \BibitemOpen
  \bibfield  {author} {\bibinfo {author} {\bibfnamefont {R.~T.}\ \bibnamefont
  {Clay}}\ and\ \bibinfo {author} {\bibfnamefont {R.~P.}\ \bibnamefont
  {Hardikar}},\ }\href {\doibase 10.1103/PhysRevLett.95.096401} {\bibfield
  {journal} {\bibinfo  {journal} {Phys. Rev. Lett.}\ }\textbf {\bibinfo
  {volume} {95}},\ \bibinfo {pages} {096401} (\bibinfo {year}
  {2005})}\BibitemShut {NoStop}%
\bibitem [{\citenamefont {Hardikar}\ and\ \citenamefont
  {Clay}(2007)}]{Hardikar2007}%
  \BibitemOpen
  \bibfield  {author} {\bibinfo {author} {\bibfnamefont {R.~P.}\ \bibnamefont
  {Hardikar}}\ and\ \bibinfo {author} {\bibfnamefont {R.~T.}\ \bibnamefont
  {Clay}},\ }\href {\doibase 10.1103/PhysRevB.75.245103} {\bibfield  {journal}
  {\bibinfo  {journal} {Phys. Rev. B}\ }\textbf {\bibinfo {volume} {75}},\
  \bibinfo {pages} {245103} (\bibinfo {year} {2007})}\BibitemShut {NoStop}%
\bibitem [{\citenamefont {Hohenadler}\ and\ \citenamefont
  {Assaad}(2013)}]{Hohenadler2013}%
  \BibitemOpen
  \bibfield  {author} {\bibinfo {author} {\bibfnamefont {M.}~\bibnamefont
  {Hohenadler}}\ and\ \bibinfo {author} {\bibfnamefont {F.~F.}\ \bibnamefont
  {Assaad}},\ }\href {\doibase 10.1103/PhysRevB.87.075149} {\bibfield
  {journal} {\bibinfo  {journal} {Phys. Rev. B}\ }\textbf {\bibinfo {volume}
  {87}},\ \bibinfo {pages} {075149} (\bibinfo {year} {2013})}\BibitemShut
  {NoStop}%
\bibitem [{\citenamefont {Sangiovanni}\ \emph {et~al.}(2005)\citenamefont
  {Sangiovanni}, \citenamefont {Capone}, \citenamefont {Castellani},\ and\
  \citenamefont {Grilli}}]{Sangiovanni2005}%
  \BibitemOpen
  \bibfield  {author} {\bibinfo {author} {\bibfnamefont {G.}~\bibnamefont
  {Sangiovanni}}, \bibinfo {author} {\bibfnamefont {M.}~\bibnamefont {Capone}},
  \bibinfo {author} {\bibfnamefont {C.}~\bibnamefont {Castellani}}, \ and\
  \bibinfo {author} {\bibfnamefont {M.}~\bibnamefont {Grilli}},\ }\href
  {\doibase 10.1103/PhysRevLett.94.026401} {\bibfield  {journal} {\bibinfo
  {journal} {Phys. Rev. Lett.}\ }\textbf {\bibinfo {volume} {94}},\ \bibinfo
  {pages} {026401} (\bibinfo {year} {2005})}\BibitemShut {NoStop}%
\bibitem [{\citenamefont {Sangiovanni}\ \emph {et~al.}(2006)\citenamefont
  {Sangiovanni}, \citenamefont {Gunnarsson}, \citenamefont {Koch},
  \citenamefont {Castellani},\ and\ \citenamefont {Capone}}]{Sangiovanni2006}%
  \BibitemOpen
  \bibfield  {author} {\bibinfo {author} {\bibfnamefont {G.}~\bibnamefont
  {Sangiovanni}}, \bibinfo {author} {\bibfnamefont {O.}~\bibnamefont
  {Gunnarsson}}, \bibinfo {author} {\bibfnamefont {E.}~\bibnamefont {Koch}},
  \bibinfo {author} {\bibfnamefont {C.}~\bibnamefont {Castellani}}, \ and\
  \bibinfo {author} {\bibfnamefont {M.}~\bibnamefont {Capone}},\ }\href
  {\doibase 10.1103/PhysRevLett.97.046404} {\bibfield  {journal} {\bibinfo
  {journal} {Phys. Rev. Lett.}\ }\textbf {\bibinfo {volume} {97}},\ \bibinfo
  {pages} {046404} (\bibinfo {year} {2006})}\BibitemShut {NoStop}%
\bibitem [{\citenamefont {Werner}\ and\ \citenamefont
  {Millis}(2007)}]{Werner2007}%
  \BibitemOpen
  \bibfield  {author} {\bibinfo {author} {\bibfnamefont {P.}~\bibnamefont
  {Werner}}\ and\ \bibinfo {author} {\bibfnamefont {A.~J.}\ \bibnamefont
  {Millis}},\ }\href {\doibase 10.1103/PhysRevLett.99.146404} {\bibfield
  {journal} {\bibinfo  {journal} {Phys. Rev. Lett.}\ }\textbf {\bibinfo
  {volume} {99}},\ \bibinfo {pages} {146404} (\bibinfo {year}
  {2007})}\BibitemShut {NoStop}%
\bibitem [{\citenamefont {Bauer}\ and\ \citenamefont
  {Hewson}(2010)}]{Bauer2010}%
  \BibitemOpen
  \bibfield  {author} {\bibinfo {author} {\bibfnamefont {J.}~\bibnamefont
  {Bauer}}\ and\ \bibinfo {author} {\bibfnamefont {A.~C.}\ \bibnamefont
  {Hewson}},\ }\href {\doibase 10.1103/PhysRevB.81.235113} {\bibfield
  {journal} {\bibinfo  {journal} {Phys. Rev. B}\ }\textbf {\bibinfo {volume}
  {81}},\ \bibinfo {pages} {235113} (\bibinfo {year} {2010})}\BibitemShut
  {NoStop}%
\bibitem [{\citenamefont {Murakami}\ \emph {et~al.}(2013)\citenamefont
  {Murakami}, \citenamefont {Werner}, \citenamefont {Tsuji},\ and\
  \citenamefont {Aoki}}]{Murakami2013}%
  \BibitemOpen
  \bibfield  {author} {\bibinfo {author} {\bibfnamefont {Y.}~\bibnamefont
  {Murakami}}, \bibinfo {author} {\bibfnamefont {P.}~\bibnamefont {Werner}},
  \bibinfo {author} {\bibfnamefont {N.}~\bibnamefont {Tsuji}}, \ and\ \bibinfo
  {author} {\bibfnamefont {H.}~\bibnamefont {Aoki}},\ }\href {\doibase
  10.1103/PhysRevB.88.125126} {\bibfield  {journal} {\bibinfo  {journal} {Phys.
  Rev. B}\ }\textbf {\bibinfo {volume} {88}},\ \bibinfo {pages} {125126}
  (\bibinfo {year} {2013})}\BibitemShut {NoStop}%
\bibitem [{\citenamefont {Wang}\ \emph {et~al.}(2012)\citenamefont {Wang},
  \citenamefont {Xiang}, \citenamefont {Wang}, \citenamefont {Wang},
  \citenamefont {Yang},\ and\ \citenamefont {Lee}}]{Wang2012}%
  \BibitemOpen
  \bibfield  {author} {\bibinfo {author} {\bibfnamefont {W.-S.}\ \bibnamefont
  {Wang}}, \bibinfo {author} {\bibfnamefont {Y.-Y.}\ \bibnamefont {Xiang}},
  \bibinfo {author} {\bibfnamefont {Q.-H.}\ \bibnamefont {Wang}}, \bibinfo
  {author} {\bibfnamefont {F.}~\bibnamefont {Wang}}, \bibinfo {author}
  {\bibfnamefont {F.}~\bibnamefont {Yang}}, \ and\ \bibinfo {author}
  {\bibfnamefont {D.-H.}\ \bibnamefont {Lee}},\ }\href {\doibase
  10.1103/PhysRevB.85.035414} {\bibfield  {journal} {\bibinfo  {journal} {Phys.
  Rev. B}\ }\textbf {\bibinfo {volume} {85}},\ \bibinfo {pages} {035414}
  (\bibinfo {year} {2012})}\BibitemShut {NoStop}%
\bibitem [{\citenamefont {Xiang}\ \emph
  {et~al.}(2012{\natexlab{a}})\citenamefont {Xiang}, \citenamefont {Wang},
  \citenamefont {Wang},\ and\ \citenamefont {Lee}}]{Xiang2012a}%
  \BibitemOpen
  \bibfield  {author} {\bibinfo {author} {\bibfnamefont {Y.-Y.}\ \bibnamefont
  {Xiang}}, \bibinfo {author} {\bibfnamefont {W.-S.}\ \bibnamefont {Wang}},
  \bibinfo {author} {\bibfnamefont {Q.-H.}\ \bibnamefont {Wang}}, \ and\
  \bibinfo {author} {\bibfnamefont {D.-H.}\ \bibnamefont {Lee}},\ }\href
  {\doibase 10.1103/PhysRevB.86.024523} {\bibfield  {journal} {\bibinfo
  {journal} {Phys. Rev. B}\ }\textbf {\bibinfo {volume} {86}},\ \bibinfo
  {pages} {024523} (\bibinfo {year} {2012}{\natexlab{a}})}\BibitemShut
  {NoStop}%
\bibitem [{\citenamefont {Nowadnick}\ \emph {et~al.}(2012)\citenamefont
  {Nowadnick}, \citenamefont {Johnston}, \citenamefont {Moritz}, \citenamefont
  {Scalettar},\ and\ \citenamefont {Devereaux}}]{Nowadnick2012}%
  \BibitemOpen
  \bibfield  {author} {\bibinfo {author} {\bibfnamefont {E.~A.}\ \bibnamefont
  {Nowadnick}}, \bibinfo {author} {\bibfnamefont {S.}~\bibnamefont {Johnston}},
  \bibinfo {author} {\bibfnamefont {B.}~\bibnamefont {Moritz}}, \bibinfo
  {author} {\bibfnamefont {R.~T.}\ \bibnamefont {Scalettar}}, \ and\ \bibinfo
  {author} {\bibfnamefont {T.~P.}\ \bibnamefont {Devereaux}},\ }\href {\doibase
  10.1103/PhysRevLett.109.246404} {\bibfield  {journal} {\bibinfo  {journal}
  {Phys. Rev. Lett.}\ }\textbf {\bibinfo {volume} {109}},\ \bibinfo {pages}
  {246404} (\bibinfo {year} {2012})}\BibitemShut {NoStop}%
\bibitem [{\citenamefont {Scalapino}\ \emph {et~al.}(1986)\citenamefont
  {Scalapino}, \citenamefont {Loh~Jr},\ and\ \citenamefont
  {Hirsch}}]{Scalapino1986}%
  \BibitemOpen
  \bibfield  {author} {\bibinfo {author} {\bibfnamefont {D.~J.}\ \bibnamefont
  {Scalapino}}, \bibinfo {author} {\bibfnamefont {E.}~\bibnamefont {Loh~Jr}}, \
  and\ \bibinfo {author} {\bibfnamefont {J.~E.}\ \bibnamefont {Hirsch}},\
  }\href {http://journals.aps.org/prb/abstract/10.1103/PhysRevB.34.8190}
  {\bibfield  {journal} {\bibinfo  {journal} {Phys. Rev. B}\ }\textbf {\bibinfo
  {volume} {34}},\ \bibinfo {pages} {8190} (\bibinfo {year}
  {1986})}\BibitemShut {NoStop}%
\bibitem [{\citenamefont {Bickers}\ \emph {et~al.}(1989)\citenamefont
  {Bickers}, \citenamefont {Scalapino},\ and\ \citenamefont
  {White}}]{Bickers1989}%
  \BibitemOpen
  \bibfield  {author} {\bibinfo {author} {\bibfnamefont {N.~E.}\ \bibnamefont
  {Bickers}}, \bibinfo {author} {\bibfnamefont {D.~J.}\ \bibnamefont
  {Scalapino}}, \ and\ \bibinfo {author} {\bibfnamefont {S.~R.}\ \bibnamefont
  {White}},\ }\href {\doibase 10.1103/PhysRevLett.62.961} {\bibfield  {journal}
  {\bibinfo  {journal} {Phys. Rev. Lett.}\ }\textbf {\bibinfo {volume} {62}},\
  \bibinfo {pages} {961} (\bibinfo {year} {1989})}\BibitemShut {NoStop}%
\bibitem [{\citenamefont {Honerkamp}\ \emph {et~al.}(2007)\citenamefont
  {Honerkamp}, \citenamefont {Fu},\ and\ \citenamefont {Lee}}]{Honerkamp2007}%
  \BibitemOpen
  \bibfield  {author} {\bibinfo {author} {\bibfnamefont {C.}~\bibnamefont
  {Honerkamp}}, \bibinfo {author} {\bibfnamefont {H.~C.}\ \bibnamefont {Fu}}, \
  and\ \bibinfo {author} {\bibfnamefont {D.-H.}\ \bibnamefont {Lee}},\ }\href
  {\doibase 10.1103/PhysRevB.75.014503} {\bibfield  {journal} {\bibinfo
  {journal} {Phys. Rev. B}\ }\textbf {\bibinfo {volume} {75}},\ \bibinfo
  {pages} {014503} (\bibinfo {year} {2007})}\BibitemShut {NoStop}%
\bibitem [{\citenamefont {Yang}\ and\ \citenamefont {Zhang}(1990)}]{Yang1990}%
  \BibitemOpen
  \bibfield  {author} {\bibinfo {author} {\bibfnamefont {C.~N.}\ \bibnamefont
  {Yang}}\ and\ \bibinfo {author} {\bibfnamefont {S.}~\bibnamefont {Zhang}},\
  }\href {\doibase 10.1142/S0217984990000933} {\bibfield  {journal} {\bibinfo
  {journal} {Mod. Phys. Lett. B}\ }\textbf {\bibinfo {volume} {04}},\ \bibinfo
  {pages} {759} (\bibinfo {year} {1990})}\BibitemShut {NoStop}%
\bibitem [{\citenamefont {Noack}\ \emph {et~al.}(1991)\citenamefont {Noack},
  \citenamefont {Scalapino},\ and\ \citenamefont {Scalettar}}]{Noack1991}%
  \BibitemOpen
  \bibfield  {author} {\bibinfo {author} {\bibfnamefont {R.~M.}\ \bibnamefont
  {Noack}}, \bibinfo {author} {\bibfnamefont {D.~J.}\ \bibnamefont
  {Scalapino}}, \ and\ \bibinfo {author} {\bibfnamefont {R.~T.}\ \bibnamefont
  {Scalettar}},\ }\href {\doibase 10.1103/PhysRevLett.66.778} {\bibfield
  {journal} {\bibinfo  {journal} {Phys. Rev. Lett.}\ }\textbf {\bibinfo
  {volume} {66}},\ \bibinfo {pages} {778} (\bibinfo {year} {1991})}\BibitemShut
  {NoStop}%
\bibitem [{\citenamefont {Macridin}\ \emph {et~al.}(2006)\citenamefont
  {Macridin}, \citenamefont {Moritz}, \citenamefont {Jarrell},\ and\
  \citenamefont {Maier}}]{Macridin2006}%
  \BibitemOpen
  \bibfield  {author} {\bibinfo {author} {\bibfnamefont {A.}~\bibnamefont
  {Macridin}}, \bibinfo {author} {\bibfnamefont {B.}~\bibnamefont {Moritz}},
  \bibinfo {author} {\bibfnamefont {M.}~\bibnamefont {Jarrell}}, \ and\
  \bibinfo {author} {\bibfnamefont {T.}~\bibnamefont {Maier}},\ }\href
  {\doibase 10.1103/PhysRevLett.97.056402} {\bibfield  {journal} {\bibinfo
  {journal} {Phys. Rev. Lett.}\ }\textbf {\bibinfo {volume} {97}},\ \bibinfo
  {pages} {056402} (\bibinfo {year} {2006})}\BibitemShut {NoStop}%
\bibitem [{\citenamefont {Pao}\ and\ \citenamefont
  {Sch眉ttler}(1998)}]{Pao1998}%
  \BibitemOpen
  \bibfield  {author} {\bibinfo {author} {\bibfnamefont {C.-H.}\ \bibnamefont
  {Pao}}\ and\ \bibinfo {author} {\bibfnamefont {H.-B.}\ \bibnamefont
  {Sch眉ttler}},\ }\href {\doibase 10.1103/PhysRevB.57.5051} {\bibfield
  {journal} {\bibinfo  {journal} {Phys. Rev. B}\ }\textbf {\bibinfo {volume}
  {57}},\ \bibinfo {pages} {5051} (\bibinfo {year} {1998})}\BibitemShut
  {NoStop}%
\bibitem [{\citenamefont {Mermin}\ and\ \citenamefont
  {Wagner}(1966)}]{Mermin1966}%
  \BibitemOpen
  \bibfield  {author} {\bibinfo {author} {\bibfnamefont {N.~D.}\ \bibnamefont
  {Mermin}}\ and\ \bibinfo {author} {\bibfnamefont {H.}~\bibnamefont
  {Wagner}},\ }\href {\doibase 10.1103/PhysRevLett.17.1133} {\bibfield
  {journal} {\bibinfo  {journal} {Phys. Rev. Lett.}\ }\textbf {\bibinfo
  {volume} {17}},\ \bibinfo {pages} {1133} (\bibinfo {year}
  {1966})}\BibitemShut {NoStop}%
\bibitem [{\citenamefont {Kosterlitz}\ and\ \citenamefont
  {Thouless}(1973)}]{Kosterlitz1973}%
  \BibitemOpen
  \bibfield  {author} {\bibinfo {author} {\bibfnamefont {J.~M.}\ \bibnamefont
  {Kosterlitz}}\ and\ \bibinfo {author} {\bibfnamefont {D.~J.}\ \bibnamefont
  {Thouless}},\ }\href {\doibase 10.1088/0022-3719/6/7/010} {\bibfield
  {journal} {\bibinfo  {journal} {J. Phys. C: Solid State Phys.}\ }\textbf
  {\bibinfo {volume} {6}},\ \bibinfo {pages} {1181} (\bibinfo {year}
  {1973})}\BibitemShut {NoStop}%
\bibitem [{\citenamefont {Qing-Yan}\ \emph {et~al.}(2012)\citenamefont
  {Qing-Yan}, \citenamefont {Zhi}, \citenamefont {Wen-Hao}, \citenamefont
  {Zuo-Cheng}, \citenamefont {Jin-Song}, \citenamefont {Wei}, \citenamefont
  {Hao}, \citenamefont {Yun-Bo}, \citenamefont {Peng}, \citenamefont {Kai}
  \emph {et~al.}}]{Qing-Yan2012}%
  \BibitemOpen
  \bibfield  {author} {\bibinfo {author} {\bibfnamefont {W.}~\bibnamefont
  {Qing-Yan}}, \bibinfo {author} {\bibfnamefont {L.}~\bibnamefont {Zhi}},
  \bibinfo {author} {\bibfnamefont {Z.}~\bibnamefont {Wen-Hao}}, \bibinfo
  {author} {\bibfnamefont {Z.}~\bibnamefont {Zuo-Cheng}}, \bibinfo {author}
  {\bibfnamefont {Z.}~\bibnamefont {Jin-Song}}, \bibinfo {author}
  {\bibfnamefont {L.}~\bibnamefont {Wei}}, \bibinfo {author} {\bibfnamefont
  {D.}~\bibnamefont {Hao}}, \bibinfo {author} {\bibfnamefont {O.}~\bibnamefont
  {Yun-Bo}}, \bibinfo {author} {\bibfnamefont {D.}~\bibnamefont {Peng}},
  \bibinfo {author} {\bibfnamefont {C.}~\bibnamefont {Kai}},  \emph {et~al.},\
  }\href@noop {} {\bibfield  {journal} {\bibinfo  {journal} {Chinese Physics
  Letters}\ }\textbf {\bibinfo {volume} {29}},\ \bibinfo {pages} {037402}
  (\bibinfo {year} {2012})}\BibitemShut {NoStop}%
\bibitem [{\citenamefont {Xiang}\ \emph
  {et~al.}(2012{\natexlab{b}})\citenamefont {Xiang}, \citenamefont {Wang},
  \citenamefont {Wang}, \citenamefont {Wang},\ and\ \citenamefont
  {Lee}}]{Xiang2012}%
  \BibitemOpen
  \bibfield  {author} {\bibinfo {author} {\bibfnamefont {Y.-Y.}\ \bibnamefont
  {Xiang}}, \bibinfo {author} {\bibfnamefont {F.}~\bibnamefont {Wang}},
  \bibinfo {author} {\bibfnamefont {D.}~\bibnamefont {Wang}}, \bibinfo {author}
  {\bibfnamefont {Q.-H.}\ \bibnamefont {Wang}}, \ and\ \bibinfo {author}
  {\bibfnamefont {D.-H.}\ \bibnamefont {Lee}},\ }\href
  {http://link.aps.org/doi/10.1103/PhysRevB.86.134508} {\bibfield  {journal}
  {\bibinfo  {journal} {Phys. Rev. B}\ }\textbf {\bibinfo {volume} {86}}
  (\bibinfo {year} {2012}{\natexlab{b}})}\BibitemShut {NoStop}%
\bibitem [{\citenamefont {Zhong}\ and\ \citenamefont
  {Sch\"uttler}(1992)}]{Zhong1992}%
  \BibitemOpen
  \bibfield  {author} {\bibinfo {author} {\bibfnamefont {J.}~\bibnamefont
  {Zhong}}\ and\ \bibinfo {author} {\bibfnamefont {H.-B.}\ \bibnamefont
  {Sch\"uttler}},\ }\href {\doibase 10.1103/PhysRevLett.69.1600} {\bibfield
  {journal} {\bibinfo  {journal} {Phys. Rev. Lett.}\ }\textbf {\bibinfo
  {volume} {69}},\ \bibinfo {pages} {1600} (\bibinfo {year}
  {1992})}\BibitemShut {NoStop}%
\bibitem [{\citenamefont {Sakai}\ \emph {et~al.}(1997)\citenamefont {Sakai},
  \citenamefont {Poilblanc},\ and\ \citenamefont {Scalapino}}]{Sakai1997}%
  \BibitemOpen
  \bibfield  {author} {\bibinfo {author} {\bibfnamefont {T.}~\bibnamefont
  {Sakai}}, \bibinfo {author} {\bibfnamefont {D.}~\bibnamefont {Poilblanc}}, \
  and\ \bibinfo {author} {\bibfnamefont {D.~J.}\ \bibnamefont {Scalapino}},\
  }\href {\doibase 10.1103/PhysRevB.55.8445} {\bibfield  {journal} {\bibinfo
  {journal} {Phys. Rev. B}\ }\textbf {\bibinfo {volume} {55}},\ \bibinfo
  {pages} {8445} (\bibinfo {year} {1997})}\BibitemShut {NoStop}%
\bibitem [{\citenamefont {Huang}\ \emph {et~al.}(2003)\citenamefont {Huang},
  \citenamefont {Hanke}, \citenamefont {Arrigoni},\ and\ \citenamefont
  {Scalapino}}]{Huang2003}%
  \BibitemOpen
  \bibfield  {author} {\bibinfo {author} {\bibfnamefont {Z.~B.}\ \bibnamefont
  {Huang}}, \bibinfo {author} {\bibfnamefont {W.}~\bibnamefont {Hanke}},
  \bibinfo {author} {\bibfnamefont {E.}~\bibnamefont {Arrigoni}}, \ and\
  \bibinfo {author} {\bibfnamefont {D.~J.}\ \bibnamefont {Scalapino}},\ }\href
  {\doibase 10.1103/PhysRevB.68.220507} {\bibfield  {journal} {\bibinfo
  {journal} {Phys. Rev. B}\ }\textbf {\bibinfo {volume} {68}},\ \bibinfo
  {pages} {220507} (\bibinfo {year} {2003})}\BibitemShut {NoStop}%
\bibitem [{\citenamefont {Mishchenko}\ and\ \citenamefont
  {Nagaosa}(2004)}]{Mishchenko2004}%
  \BibitemOpen
  \bibfield  {author} {\bibinfo {author} {\bibfnamefont {A.~S.}\ \bibnamefont
  {Mishchenko}}\ and\ \bibinfo {author} {\bibfnamefont {N.}~\bibnamefont
  {Nagaosa}},\ }\href {\doibase 10.1103/PhysRevLett.93.036402} {\bibfield
  {journal} {\bibinfo  {journal} {Phys. Rev. Lett.}\ }\textbf {\bibinfo
  {volume} {93}},\ \bibinfo {pages} {036402} (\bibinfo {year}
  {2004})}\BibitemShut {NoStop}%
\bibitem [{\citenamefont {Macridin}\ \emph {et~al.}(2004)\citenamefont
  {Macridin}, \citenamefont {Sawatzky},\ and\ \citenamefont
  {Jarrell}}]{Macridin2004}%
  \BibitemOpen
  \bibfield  {author} {\bibinfo {author} {\bibfnamefont {A.}~\bibnamefont
  {Macridin}}, \bibinfo {author} {\bibfnamefont {G.~A.}\ \bibnamefont
  {Sawatzky}}, \ and\ \bibinfo {author} {\bibfnamefont {M.}~\bibnamefont
  {Jarrell}},\ }\href {\doibase 10.1103/PhysRevB.69.245111} {\bibfield
  {journal} {\bibinfo  {journal} {Phys. Rev. B}\ }\textbf {\bibinfo {volume}
  {69}},\ \bibinfo {pages} {245111} (\bibinfo {year} {2004})}\BibitemShut
  {NoStop}%
\bibitem [{\citenamefont {Huang}\ \emph {et~al.}(2011)\citenamefont {Huang},
  \citenamefont {Lin},\ and\ \citenamefont {Arrigoni}}]{Huang2011}%
  \BibitemOpen
  \bibfield  {author} {\bibinfo {author} {\bibfnamefont {Z.~B.}\ \bibnamefont
  {Huang}}, \bibinfo {author} {\bibfnamefont {H.~Q.}\ \bibnamefont {Lin}}, \
  and\ \bibinfo {author} {\bibfnamefont {E.}~\bibnamefont {Arrigoni}},\ }\href
  {\doibase 10.1103/PhysRevB.83.064521} {\bibfield  {journal} {\bibinfo
  {journal} {Phys. Rev. B}\ }\textbf {\bibinfo {volume} {83}},\ \bibinfo
  {pages} {064521} (\bibinfo {year} {2011})}\BibitemShut {NoStop}%
\bibitem [{\citenamefont {Macridin}\ \emph {et~al.}(2012)\citenamefont
  {Macridin}, \citenamefont {Moritz}, \citenamefont {Jarrell},\ and\
  \citenamefont {Maier}}]{Macridin2012}%
  \BibitemOpen
  \bibfield  {author} {\bibinfo {author} {\bibfnamefont {A.}~\bibnamefont
  {Macridin}}, \bibinfo {author} {\bibfnamefont {B.}~\bibnamefont {Moritz}},
  \bibinfo {author} {\bibfnamefont {M.}~\bibnamefont {Jarrell}}, \ and\
  \bibinfo {author} {\bibfnamefont {T.}~\bibnamefont {Maier}},\ }\href
  {\doibase 10.1088/0953-8984/24/47/475603} {\bibfield  {journal} {\bibinfo
  {journal} {J. Phys.: Condens. Matter}\ }\textbf {\bibinfo {volume} {24}},\
  \bibinfo {pages} {475603} (\bibinfo {year} {2012})}\BibitemShut {NoStop}%
\bibitem [{\citenamefont {Wang}\ \emph {et~al.}(2014)\citenamefont {Wang},
  \citenamefont {Yang},\ and\ \citenamefont {Wang}}]{Wang2014}%
  \BibitemOpen
  \bibfield  {author} {\bibinfo {author} {\bibfnamefont {W.-S.}\ \bibnamefont
  {Wang}}, \bibinfo {author} {\bibfnamefont {Y.}~\bibnamefont {Yang}}, \ and\
  \bibinfo {author} {\bibfnamefont {Q.-H.}\ \bibnamefont {Wang}},\ }\href
  {\doibase 10.1103/PhysRevB.90.094514} {\bibfield  {journal} {\bibinfo
  {journal} {Phys. Rev. B}\ }\textbf {\bibinfo {volume} {90}},\ \bibinfo
  {pages} {094514} (\bibinfo {year} {2014})}\BibitemShut {NoStop}%
\bibitem [{\citenamefont {Honerkamp}\ \emph {et~al.}(2001)\citenamefont
  {Honerkamp}, \citenamefont {Salmhofer}, \citenamefont {Furukawa},\ and\
  \citenamefont {Rice}}]{Honerkamp2001}%
  \BibitemOpen
  \bibfield  {author} {\bibinfo {author} {\bibfnamefont {C.}~\bibnamefont
  {Honerkamp}}, \bibinfo {author} {\bibfnamefont {M.}~\bibnamefont
  {Salmhofer}}, \bibinfo {author} {\bibfnamefont {N.}~\bibnamefont {Furukawa}},
  \ and\ \bibinfo {author} {\bibfnamefont {T.}~\bibnamefont {Rice}},\ }\href
  {http://link.aps.org/doi/10.1103/PhysRevB.63.035109} {\bibfield  {journal}
  {\bibinfo  {journal} {Phys. Rev. B}\ }\textbf {\bibinfo {volume} {63}}
  (\bibinfo {year} {2001})}\BibitemShut {NoStop}%
\bibitem [{\citenamefont {Metzner}\ \emph {et~al.}(2012)\citenamefont
  {Metzner}, \citenamefont {Salmhofer}, \citenamefont {Honerkamp},
  \citenamefont {Meden},\ and\ \citenamefont {Sch枚nhammer}}]{Metzner2012}%
  \BibitemOpen
  \bibfield  {author} {\bibinfo {author} {\bibfnamefont {W.}~\bibnamefont
  {Metzner}}, \bibinfo {author} {\bibfnamefont {M.}~\bibnamefont {Salmhofer}},
  \bibinfo {author} {\bibfnamefont {C.}~\bibnamefont {Honerkamp}}, \bibinfo
  {author} {\bibfnamefont {V.}~\bibnamefont {Meden}}, \ and\ \bibinfo {author}
  {\bibfnamefont {K.}~\bibnamefont {Sch枚nhammer}},\ }\href {\doibase
  10.1103/RevModPhys.84.299} {\bibfield  {journal} {\bibinfo  {journal} {Rev.
  Mod. Phys.}\ }\textbf {\bibinfo {volume} {84}},\ \bibinfo {pages} {299}
  (\bibinfo {year} {2012})}\BibitemShut {NoStop}%
\bibitem [{\citenamefont {Platt}\ \emph {et~al.}(2013)\citenamefont {Platt},
  \citenamefont {Hanke},\ and\ \citenamefont {Thomale}}]{Platt2013}%
  \BibitemOpen
  \bibfield  {author} {\bibinfo {author} {\bibfnamefont {C.}~\bibnamefont
  {Platt}}, \bibinfo {author} {\bibfnamefont {W.}~\bibnamefont {Hanke}}, \ and\
  \bibinfo {author} {\bibfnamefont {R.}~\bibnamefont {Thomale}},\ }\href
  {\doibase 10.1080/00018732.2013.862020} {\bibfield  {journal} {\bibinfo
  {journal} {Adv. Phys.}\ }\textbf {\bibinfo {volume} {62}},\ \bibinfo {pages}
  {453} (\bibinfo {year} {2013})}\BibitemShut {NoStop}%
\end{thebibliography}%

\newpage
\beginsupplement
\maketitle

\begin{center}
\textbf{\large Supplemental Material for ``Phonon enhancement of electronic orders and negative isotope effect in the Hubbard-Holstein Model on a square lattice''}
\end{center}


\section{Technical details of SM-FRG}

\begin{figure}
\includegraphics[width=0.45\textwidth]{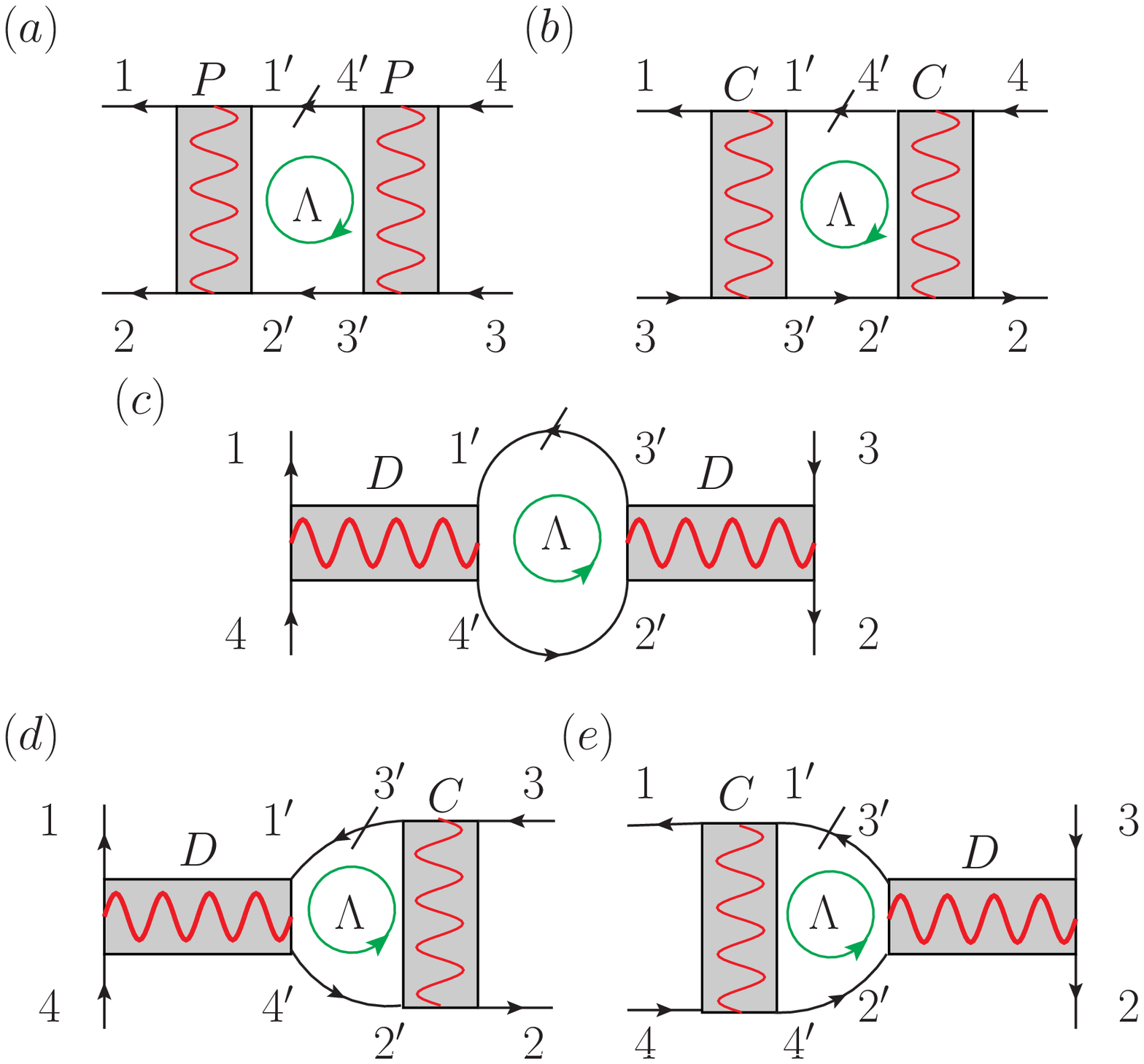}
\caption{One-loop contributions to $\partial V/\partial\Lambda$. The greyed bar and wavy line denote $V$ and $\Pi$, respectively. They add up where overlayed. Spin is conserved during fermion propagation and is left implicit. The slash denotes the single-scale propagator and can be put on either one of the fermion lines within the loop. The directed-circle indicates circulation of frequency along the loop, and $\Lambda$ the running scale.  The thin (thick) wavy line shares the loop frequency (is at zero frequency). Each diagram can be viewed as a convolution of aliases of $V$ (together with $\Pi$) via $V_{1234}=P_{(12)(43)}=C_{(13)(42)}=D_{(14)(32)}$.}
\label{fig:frgscheme}
\end{figure}

Consider the interaction hamiltonian $H_I=(1/2)c_{1\si}^\dagger c_{2\si'}^\dagger V_{1234} c_{3\si'} c_{4\si}$. Here the numerical index labels momentum/position, and we leave implicit the momentum conservation/translation symmetry. The spin SU(2) symmetry is guaranteed in the above convention for $H_I$. The idea of FRG is to get the one-particle-irreducible interaction vertex for fermions whose energy/frequency is above a scale $\La$. Equivalently, such an effective interaction is what's called pseudo-potential for fermions whose energy/frequency is below $\La$. Starting from the local $U$ at $\La=\infty$, the contributions to $\p V/\p \La$ are illustrated in \Fig{fig:frgscheme}. In principle there will also be self-energy correction to fermions, which we ignore as usual, given the fact that we are just looking for the instability of the normal state. To proceed, it is useful to define matrix aliases of the rank-4 `tensor' $V$ via
\eqa V_{1234}=P_{(12)(43)}=C_{(13)(42)}=D_{(14)(32)}.\eea
Then $\p V/\p \La$ can be compactly written as
\eqa \frac{\p V_{1234}}{\p\La} = &&[\cD\chi^{ph}(\cD-\cC)+(\cD-\cC)\chi^{ph}\cD]_{(14)(32)}\nn
                                 &&+ [\cP\chi^{pp}\cP]_{(12)(43)} - [\cC\chi^{ph}\cC]_{(13)(42)},
\label{Eq:dV}
\eea
where matrix convolutions are understood within the square brackets, and
\eqa && \cP = P +\Pi_\La, \ \ \cC = C + \Pi_\La, \ \ \cD = D + \Pi_0,\nn
     && \chi^{pp}_{(ab)(cd)} = \frac{1}{2\pi}[G_{ac}(\La)G_{bd}(-\La)+(\La\ra -\La)],\nn
     && \chi^{ph}_{(ab)(cd)} = -\frac{1}{2\pi}[G_{ac}(\La)G_{db}(\La)+(\La\ra -\La)],
\label{Eq:def}
\eea
where $\Pi$ enters as a matrix (local in real space and flat in momentum space), $G$ is the normal state Green's function, and we used a hard-cutoff in the continuous Matsubara frequency. Notice that $\Pi_0$ enters $\cD$ because the EPC induced interaction is direct in the charge channel. This is also evident from \Fig{fig:frgscheme}. Since the external lines are set at zero frequency (the frequency dependence is irrelevant for 4-point interactions in the RG sense), the frequency on the phonon lines (thickened wavy lines) overlayed by $D$ is automatically zero in \Fig{fig:frgscheme}(c)-(e).

The integration of $\p V/\p \La$ toward decreasing $\La$ generates all one-particle-irreducible corrections to $V$ from $U$ and $\Pi$ to arbitrary orders and in all possible ways. We extract from $V$ and $\Pi$ the effective interactions in the general SC/SDW/CDW channels
\eqa (V_{SC},V_{SDW},V_{CDW}) = (\cP, -\cC, 2\cD-\cC).\eea (This expression is exactly equivalent to \Eq{eq:VX} in the main text.) Since they all originate from $V+\Pi$, they are overlapped but are naturally treated on equal footing. Viewed as scattering amplitude of composite bosons, the effective interactions can be decomposed into eigen modes. For example, in the SC channel (with a zero collective momentum),
\eqa
[V_{SC}]_{(\v k,-\v k)(\v k',-\v k')} = \sum_m f_m(\v k)S_m f_m^{*}(\v k'),
\eea
where $S_m$ is the eigenvalue, and $f_m(\v k)$ is the eigenfunction. We look for the most negative eigenvalue, say $S=\min[S_m]$, with an associated eigenfunction $f(\v k)$. If $S$ diverges at a scale $\La_c$, it signals the instability of the normal state toward a SC state, with a pairing function described by $f(\v k)$. Similar analysis can be performed in the CDW/SDW channels, with the only exception that in general the collective momentum $\v q$ in such channels is nonzero. Since $\v q$ is a good quantum number in the respective channels, one performs the mode decomposition at each $\v q$. There are multiple modes at each $\v q$, but we are interested in the globally leading mode among all $\v q$. In this way one determines both the ordering vector $\v Q$ and the structure of the order parameter by the leading eigenfunction.
Finally, the instability channel is determined by comparing the leading eigenvalues in the CDW/SDW/SC channels.

In principle, the above procedure is able to capture the most general candidate order parameters. In practice, however, it is impossible to keep all elements of $V$ for computation. Fortunately, the order parameters are always local or short-ranged. This is notwithstanding the possible long-range correlations between the order parameters. For example, the s-wave pairing in the BCS theory is local, since the gap function is a constant in momentum space. The order parameter in usual Landau theories are assumed to be local. The d-wave pairing is nonlocal but short-ranged. The usual CDW/SDW orders are ordering of site-local charges/spins. The valence-bond order is on-bond but short-ranged. In fact, if the order parameter is very nonlocal, it is not likely to be stable. The idea is, if it is not an instability at the tree level, it has to be induced by the overlapping channel. But if the induced order parameter is very nonlocal, it must be true that the donor channel has already developed long-range fluctuations and is ready to order first. These considerations suggest that most elements of the `tensor' $V$ are irrelevant in the RG sense and can be truncated. \Eq{Eq:dV} suggests how this can be done. For fermions, all 4-point interactions are marginal in the RG sense, and the only way a marginal operator could become relevant is through coherent and repeated scattering in a particular channel. Therefore, it is sufficient to truncate the range between 1 and 2, between 3 and 4, in  $\cP_{(12)(43)}$, but leaving the range between the two groups arbitrary (thus thermodynamical limit is not spoiled). Similar considerations apply to $\cC$ and $\cD$. Eventually the same type of truncations can be applied in the effective interactions $V_{CDW/SDW/SC}$. Such truncations keep the potentially singular contributions in all channels and their overlaps, underlying the key idea of the SM-FRG.~\cite{Wang2012,Xiang2012a,Wang2014} The merit of SM-FRG is: 1) It guarantees hermiticity of the truncated interactions; 2) It is asymptotically exact if the truncation range is enlarged; 3) It respects all underlying symmetries, and in particular it respects momentum conservation exactly. 4) In systems with multi-orbitals or complex unitcell, it is important to keep the momentum dependence of the Bloch states, both radial and tangential to the Fermi surface. This is guaranteed in SM-FRG since it works with Green's functions in the orbital basis. We take these as advantages of SM-FRG as compared to the patch-FRG applied in the literature.~\cite{Honerkamp2001,Metzner2012,Platt2013}

{\em BCS limit}: We notice that if only \Fig{fig:frgscheme}(a), the pairing channel, is kept, the BCS theory is trivially reproduced. For this to be valid, one requires $\La_c\ll \w_D\ll W$ and the absence of any nesting, so that the contributions from the other channels, \Fig{fig:frgscheme}(b)-(e), are negligible. To make analytical solution accessible, we approximate $\Pi_\nu$ as a step function, $\Pi_\nu =-\la W\theta(\w_D-|\nu|)$. Thus $\Pi_\La =0$ for $\La >\w_D$, and the RG flow above $\w_D$ merely generates a renormalized Coulomb interaction $V^*$. The flow for $\La<\w_D$ is, with $\Pi_\La=-\la W$ in the above approximation,
\eqa \p (V-\la W)/\p \La = (\rho/\La)(V-\la W)^2,\eea
where $\rho$ is the normal state density of states, and we assumed that $V_{\v k,-\v k,-\v k',\v k'}$ is independent of $\v k$ and $\v k'$, as assumed in the BCS theory. (This means that we are treating the s-wave pairing channel.) The solution is, given the boundary condition at $\La=\w_D$,
\eqa V-\la W = \frac{V^*-\la W}{1 +(\la-\mu^*)\ln(\La/\w_D)},\eea where we used $\mu^*=\rho V^*$ and $\rho W\sim 1$. There is a divergence $V-\la W\ra -\infty$ if and only if $\la-\mu^*>0$ (i.e., EPC mediated attraction overwhelms the repulsive $V^*$), at the scale
\eqa \La_c = \w_D e^{-1/(\la-\mu^*)}.\eea This is already in nice agreement with the $T_c$ in the Eliashberg theory, given the approximations in $\Pi_\nu$. This example shows that the idea of pseudopotential can be pushed down to any energy scale (not just at $\w_D$ as in the BCS theory) until it diverges, and the divergence scale is just a representative of the transition temperature $T_c$. If sufficiently strong, the CDW/SDW channels neglected in the BCS theory will clearly invalidate the latter, as revealed in the main text.\\

{\em Local limit}: On the other hand, if only the local elements of $V$ is kept, we have $V=P=C=D$. Furthermore, in the presence of particle-hole symmetry (at half filling in HHM), the second line of \Eq{Eq:dV} cancels out (in the local limit), leaving,
\eqa \frac{\p V}{\p\La} = - \frac{2}{\pi}(\Pi_0-\Pi_\La)\frac{\p\chi_\La}{\p\La}(V+\Pi_0),\eea
where $\chi_\La\sim \al/\La$ is a local susceptibility at the scale $\La$, with a factor $\al$ of order unity. This can be solved analytically,
\eqa V+\Pi_0 \sim (U+\Pi_0)\exp\left[\frac{\al\la W}{\w_D}(1-\frac{2}{\pi}\tan^{-1}\frac{\La}{\w_D}) \right],\eea where we used $\Pi_0=-\la W$. This is \Eq{Eq:Anal} in the main text.\\

\section{Polaronic band narrowing}

The coupling between electron and phonon can be formally decoupled by the Lang-Firsov transformation.
Define a unitary matrix $\cU =\exp[(\eta/\w_D)\sum_i n_i(b_i-b_i^\dag)]$,
where we recall that $\eta=g/\sqrt{2M\w_D}=\sqrt{\la W\w_D/2}$. It is easy to show that $\cH = \cU H \cU^\dag$ becomes
\eqa
\cH=&&-t\sum_{\<ij\>\si}(\t c_{i\sigma}^\dag \t c_{j\sigma}+{\rm h.c.}) -\mu\sum_{i\sigma}n_{i\sigma} +\omega_D\sum_i b_i^\dag b_i\nn && + \t U \sum_{i}(n_{i\ua}-1/2)(n_{i\da}-1/2) ,
\eea
where $\t c_i = c_i e^{-(\eta/\w_D)(b_i-b_i^\dag)}$ and $\t U = U-\la W$. The hopping part averaged over the phonon ensemble leads to a polaronic renormalization of $t\ra z t$, with
\eqa z= \exp\left[-\frac{\la W}{2\w_D}\frac{1+e^{\bt\w_D}}{e^{\beta \w_D}-1}\right].\eea This factor describes the coherent part of the kinetic energy in the presence of EPC, namely the renormalization factor for the cohorent bandwidth. For the electrons to hop coherently, one requires $T\ll \w_D$ so that phonon excitations are rare, the condition for $z$ to make sense.

\end{document}